\documentclass[fleqn,usenatbib]{mnras}
\pdfoutput=1
\usepackage{newtxtext,newtxmath}
\usepackage[T1]{fontenc}
\usepackage{ae,aecompl}
\usepackage{natbib,textcomp,placeins,epsfig,graphicx,color,amsmath,upgreek,listings,epstopdf}
\usepackage[]{threeparttable}

\usepackage{tikz}

\newcommand{\logg}{$\log g$} 
\newcommand{\teff}{$T_{\mbox{\footnotesize eff}}$}

 \newcommand{\z}{$|z|$}
\newcommand{\Rgal}{$R_{\rm Gal}$}
\newcommand{\Gaia}{{\it Gaia}\,}

\defcitealias{Myeong2019}{M19}
\defcitealias{Feuillet2020}{F20}

\title[Characterizing Gaia-Sausage-Enceladus and Sequoia]{Selecting accreted populations: metallicity, elemental abundances, and ages of the {\it Gaia}-Sausage-Enceladus and Sequoia populations}

\author[D. K. Feuillet et al.]{Diane K. Feuillet,$^{1}$\thanks{email: diane.feuillet@astro.lu.se} 
Christian L. Sahlholdt,$^{1}$
Sofia Feltzing,$^{1}$
Luca Casagrande$^{2,3}$
\\
$^{1}$Lund Observatory, Department of Astronomy and Theoretical Physics, Box 43, SE-221\,00 Lund, Sweden\\
$^{2}$Research School of Astronomy and Astrophysics, Mount Stromlo Observatory, The Australian National University, ACT 2611, Australia\\
$^{3}$ARC Centre of Excellence for All Sky Astrophysics in 3 Dimensions (ASTRO 3D), Australia \\
}

\date{Accepted XXX. Received YYY; in original form ZZZ}

\pubyear{2021}

\begin{document}
\label{firstpage}
\pagerange{\pageref{firstpage}--\pageref{lastpage}}
\maketitle

\begin{abstract}

Identifying stars found in the Milky Way as having formed {\it in situ} or accreted can be a complex and uncertain undertaking. We use \Gaia kinematics and APOGEE elemental abundances to select stars belonging to the {\it Gaia}-Sausage-Enceladus (GSE) and Sequoia accretion events. These samples are used to characterize the GSE and Sequoia population metallicity distribution functions, elemental abundance patterns, age distributions, and progenitor masses. We find that the GSE population has a mean [Fe/H] $\sim -1.15$ and a mean age of $10-12$ Gyr. GSE has a single sequence in [Mg/Fe] vs [Fe/H] consistent with the onset of SN Ia Fe contributions and uniformly low [Al/Fe] of $\sim -0.25$ dex. The derived properties of the Sequoia population are strongly dependent on the kinematic selection. We argue the selection with the least contamination is $J_{\phi}/J_{\mbox{tot}} < -0.6$ and $(J_z - J_R)/J_{\mbox{tot}} < 0.1$. This results in a mean [Fe/H] $\sim -1.3$ and a mean age of $12-14$ Gyr. The Sequoia population has a complex elemental abundance distribution with mainly high [Mg/Fe] stars.
We use the GSE [Al/Fe] vs [Mg/H] abundance distribution to inform a chemically-based selection of accreted stars, which is used to remove possible contaminant stars from the GSE and Sequoia samples.

\end{abstract}


\section{Introduction}

The halo of the Milky Way is known to host several accreted stellar populations. Recent estimates find that as much as $40-80\%$ of the stellar halo mass originates in accreted galaxies \citep{Mackereth2020, Naidu2020}. The fraction of accreted matter in the halo depends on the Galactocentric radius. Accreted stars can be identified in several ways. Many accreted stellar populations can be identified in phase-space \citep[e.g.][]{Koppelman2019, Naidu2021}. Elemental abundance patterns can also be used to identify accreted populations because stars born together have similar abundance patterns \citep[see][]{Freeman2002, Nissen2010, Hayes2018}. In addition, local dwarf galaxies are known to have elemental abundance patterns that are distinct from those found in the Milky Way (\citealt{Venn2004}; \citealt*{Tolstoy2009}). Pre-\Gaia efforts to identify accreted stars were limited to stellar populations that had experienced minimal disruption and could therefore be visually associated either in phase-space or on the sky \citep{Helmi1999, Belokurov2006}. However, elemental abundance measurements of nearby dwarf stars with halo kinematics suggested a significant accreted population was present \citep{Nissen2010}. Identifying and characterizing these accreted populations and their progenitor galaxies provides a glimpse of the Milky Way's history and give constraints for the assembly history of galaxies \citep[e.g.][]{Mcconnachie2009, Cooper2010, Deason2013}.

With the astrometric measurements of over a billion stars from the \Gaia mission \citep{Gaia2018a} has come the discovery of the {\it Gaia}-Sausage-Enceladus (GSE) accretion event \citep{Belokurov2018, Helmi2018} as well as several other accreted populations \citep[see][]{Koppelman2019, Naidu2020, Kruijssen2020, Malhan2021}, including the Sequoia accretion event identified by \citet{Myeong2019}. Significant efforts have been made to characterize these accreted populations, primarily using kinematic selections, and their progenitor galaxies. In this paper, we focus on characterizing the kinematic properties and elemental abundance patterns of the GSE and Sequoia populations in order to determine the mass of the progenitors and the age of their stellar populations with the aim to further our understanding of the Milky Way's assembly history. 

The GSE population has been the focus of many studies since its discovery, however, there remains some uncertainty about how best to select population members and therefore its properties, such as mean metallicity, remain uncertain. \citet{Belokurov2018} identified an elongated kinematic structure narrowly centered around zero velocity in the azimuthal direction with a large range of velocities in the radial direction that is considered a signature of the accreted population. \citet{Helmi2018} identified a general overdensity of stars with low to retrograde angular momentum and high orbital energy, which also have an [$\alpha$/Fe] abundance pattern consistent with an accreted origin and a mean metallicity of $-1.5$ dex. 
\citet{Koppelman2019} identified several distinct populations in the retrograde halo using a clustering algorithm and constrained GSE members to $L_Z \sim 0$, finding a metallicity distribution in agreement with \citet{Helmi2018}. \citet[]{Feuillet2020} defined a GSE selection based on stellar metallicities and found stars with $L_Z \sim 0$ and high radial action have a narrow metallicity distribution function centered at [Fe/H] $= -1.17$, suggestive of a single population. \citet{Conroy2019} found highly radial halo stars have a mean [Fe/H] $\sim -1.2$, in agreement with the kinematics and metallicity found by \citet{Feuillet2020}. Many studies of subsamples of halo stars that likely include, but are not selected to be exclusively, GSE stars find mean metallicities between $-1.0$ and $-1.5$ dex (\citealt{Haywood2018}; \citealt{Hayes2018}; \citealt*{Sahlholdt2019}; \citealt{Gallart2019}; \citealt{Conroy2019}; \citealt{Mackereth2019}; \citealt*{Amarante2020}). 
\citet{Naidu2020} detailed several of the stellar structures now known in the halo, including the GSE population, which they limited to high eccentricity stars and found the metallicity distribution function to be well fit by a simple chemical evolution model with a mean [Fe/H] $= -1.15$, also in agreement with \citet{Feuillet2020}. 

The mass of the GSE progenitor has been estimated using several methods. \citet{Belokurov2018} used cosmological simulations to constrain the virial mass to be less than $10^{10}$ M$_{\odot}$. \citet{Mackereth2019} constrained the stellar mass to be $3-10 \times 10^8$ M$_{\odot}$ by comparing dynamics and elemental abundances of accreted stars identified in APOGEE with EAGLE simulations. \citet{Mackereth2020} refined this estimate to be $3 \times 10^8$ M$_{\odot}$ for GSE using an analytical model of the Milky Way and APOGEE data. Mass estimates using mass-metallicity relations \citep{Feuillet2020, Kruijssen2020, Naidu2020} or chemical evolution modeling \citep{Helmi2018, FernandezAlvar2018, Vincenzo2019} range from $4 \times 10^8$ to $7 \times 10^9$ M$_{\odot}$ for the stellar mass of GSE.

 \citet{Myeong2019} found that several of the metal-poor, retrograde, halo globular clusters are likely accreted from a single progenitor, which they name the Sequoia. They find the Sequoia population has a mean [Fe/H] of $-1.5$ dex. \citet*{Matsuno2019} find a similar metallicity for high energy retrograde stars, [Fe/H] $= -1.6$. \citet{Monty2020} found kinematic evidence for two groups within the Sequoia population, which combined have a metallicity distribution that agrees with \citet{Myeong2019}.
\citet{Naidu2020} found that the high energy, retrograde stars typically assigned to Sequoia have three peaks in the MDF, which they assigned to three different populations, Arjuna, Sequoia, and I'itoi, and argued that the most metal-rich population may be a retrograde tail of GSE. 

The stellar mass of the Sequoia progenitor is estimated to be approximately an order of magnitude lower than that of GSE. \citet{Myeong2019} estimated the Sequoia mass to be $1.7 \times 10^8$ M$_{\odot}$ using the mass-metallicity-redshift relation of \citet{Ma2016}. \citet{Kruijssen2020} estimated the Sequoia progenitor to be $0.8 \times 10^8$ M$_{\odot}$ using a neutral network to compare cosmological simulations with the distribution of globular clusters associated with various accreted populations. By assuming a relation between halo mass and total number of globular clusters, \citet{Forbes2020} also estimate a stellar mass of $0.8 \times 10^8$ M$_{\odot}$ for the Sequoia.

The age of the GSE and Sequoia populations has not been robustly measured. The age of the blue sequence of the dual halo sequence \citep{Gaia2018b} has been estimated using isochrone fitting to be older than $11$ Gyr \citep{Gallart2019, Sahlholdt2019}. The accreted halo has been estimated to be older than $\sim 10$ Gyr by \citet{Bonaca2020} and older than $8$ Gyr by \citet*{Das2020}. \citet{Dimatteo2019} found that halo stars in APOGEE with low-[Mg/Fe] and kinematic similarities to the GSE population have ages of $9-11$ Gyr. In a sample of stars with asteroseismic measurements, \citet{Montalban2021} find seven stars with low-[Mg/Fe] and high eccentricity, consistent with the GSE population, which have a mean age of $10$ Gyr. The GSE and Sequoia progenitors are estimated to have first formed $13.5$ and $13.3$ Gyr ago, respectively, by \citet{Forbes2020} using the age-metallicity relation of globular clusters associated with each population. \citet{Vincenzo2019} modeled the chemical evolution of GSE and predicted the mean stellar age to be $\sim 12$ Gyr.

While it is hoped that strong chemical tagging, as described in \citet*{Bland-Hawthorn2010}, can be used to identify stars that formed in the same cluster with extremely high-precision elemental abundance measurements, weak chemical tagging has already been used with current spectroscopic surveys to identify stars that formed at the same Galactocentric radius or in the same type of stellar system \citep[e.g.][]{Minchev2018, Hasselquist2019}. Elemental abundance measurements of stars in local dwarf galaxies show that these systems are chemically distinct from Milky Way disc stars and often from other low-mass systems \citep[e.g.][]{Shetrone2003, Venn2004, Tolstoy2009, Hasselquist2017, Nidever2020}.

Certain elements have been found to have more diagnostic power than others when searching for accreted stars. Elemental abundance measurements in stars belonging to nearby or recently accreted dwarf galaxies can provide a prediction of the possible elemental abundance patterns expected in accreted stellar populations as compared to the Milky Way stellar populations. The nearby Magellanic Clouds have been found to be deficient in $\alpha$-elements relative with the Milky Way \citep{Lapenna2012, Mucciarelli2014, Nidever2020}. \citet{Hasselquist2017} found the recently accreted Sagittarius dwarf spheroidal galaxy to be deficient in 14 elements relative to the Milky Way using APOGEE data and confirmed these elemental abundance patterns are also present in stars belonging to the Sagittarius stream in follow-up work \citep{Hasselquist2019}.

A population of stars belonging to the halo with low $\alpha$-element and [Ni/Fe] abundances has been explored as a possible accreted population \citep[see][]{Nissen2010} and found to be deficient not only in $\alpha$-elements and Ni, but also in C+N and Al by \citet{Hayes2018}. 
\citet{Das2020} used the [Mg/Mn] vs [Al/Fe] chemical plane as suggested by \citet{Hawkins2015} to select likely accreted stars from the Milky Way halo using APOGEE DR14 data. They found these accreted stars to be between $8$ and $13$ Gyr old. 
The patterns of neutron capture elements of stars in nearby dwarf galaxies are known to be distinct from the main Milky Way stellar populations \citep[e.g.][]{Tolstoy2009, Reichert2020}. However, neutron capture  elements are not available for a majority of the APOGEE dataset, which is used in this study \citep[see exceptions in][]{Hasselquist2016, Cunha2017}.

A few studies have now inspected the elemental abundances of stars belonging to the GSE and Sequoia populations. \citet{Myeong2019} found the two populations to have lower Mg and Al abundances than the Milky Way, and to be distinct from each other. Samples of low [Mg/Fe], [Al/Fe] halo stars \citep[such as those found by][]{Nissen2010, Hayes2018, Hasselquist2019} probably contain stars belonging to GSE, Sequoia, or both. 

In this paper, we use the kinematics and elemental abundances of stars observed with \Gaia and APOGEE to determine selection criteria for the GSE and Sequoia populations (\S\ref{sec:select}). With these samples we characterize the metallicity, elemental abundances, and age distribution of both populations (\S \ref{sec:abund}) and further refine their selection (\S\ref{sec:accrete}). We discuss our findings and their implications in \S\ref{sec:discussion}.

\section{Data}
\label{sec:data}

In this study we use data from the \Gaia Data Release 2 (\Gaia DR2) \citep{Gaia2018a} and the APOGEE survey \citep{Majewski2017} Data Release 16 \citep[DR16][]{Ahumada2020, Jonsson2020}. All stars are required to have parallax measurement relative uncertainties $< 20\%$. We are interested in selecting stars in various kinematic spaces, therefore we define two datasets for our analyses: 1) stars in the \Gaia DR2 Radial Velocity Spectrometer (RVS) catalogue and 2) a crossmatch between APOGEE DR16 and \Gaia DR2. 

Radial velocity measurements and distance estimates are needed for each sample to calculate kinematic properties. The \Gaia DR2 RVS sample provides the larger dataset with full kinematic measurements down to $G \sim 15$. For this dataset we use the \Gaia DR2 RVS radial velocity measurements \citep{Gaia2018rvs}, distances from \citet*{Schonrich2019}, and \Gaia DR2 astrometry. The APOGEE DR16 + \Gaia DR2 sample provides a smaller sample for which detailed elemental abundances are available down to $H \sim 14$, which roughly corresponds to $G \sim 16$ for the giant stars used in this study. For this dataset we use the APOGEE radial velocity measurements, distances from \citet{BailerJones2018}, and \Gaia DR2 astrometry. 
The full space velocities, actions, and orbital energies were calculated by us with {\it galpy} using the `MWPotential2014' Galactic potential \citep{Bovy2013, Bovy2015}. Our study uses velocity in Galactocentric cylindrical coordinates, tangential velocity ($V_{\phi}$), radial velocity ($V_R$), and velocity in the $z$ direction ($V_z$). We also make use of the integrals of motion, focusing on angular momentum in the $z$ direction ($L_z$), radial action ($\sqrt{J_R}$), and total orbit energy ($E_n$).


Additional quality selection criteria are needed for the APOGEE DR16 + \Gaia DR2 sample to select stars with the highest precision in the elemental abundance measurements \citep[see][]{Jonsson2020}. In particular, we limit our study to giant stars in order to avoid stars forming the [Mg/Fe] `belly' noted by \citet{Jonsson2020}. We also remove stars with any STARFLAG or ASPCAPFLAG flags set, which indicate possible issues with the data reduction and spectral analysis \citep{Jonsson2020}.
The quality cuts imposed are listed in Table \ref{tab:quality}.
In addition to these quality cuts, we also remove stars in fields targeted for known clusters or accreted populations. This is done on each of the GSE and Sequoia kinematically selected samples. Examples of removed fields include those focused on Sagittarius and the Magellanic Clouds. A full list of the fields removed from the final dataset is give in Table \ref{tab:fields}. Although cluster members could in principle be removed from the RVS sample as well, only subgiants are used to determine age and our RVS sample is too bright to contain subgiant cluster members, see Section \ref{sec:ages}.



\begin{figure*}
\includegraphics[clip,width=0.67\hsize,angle=0,trim=0.5cm 5cm 2cm 5cm] {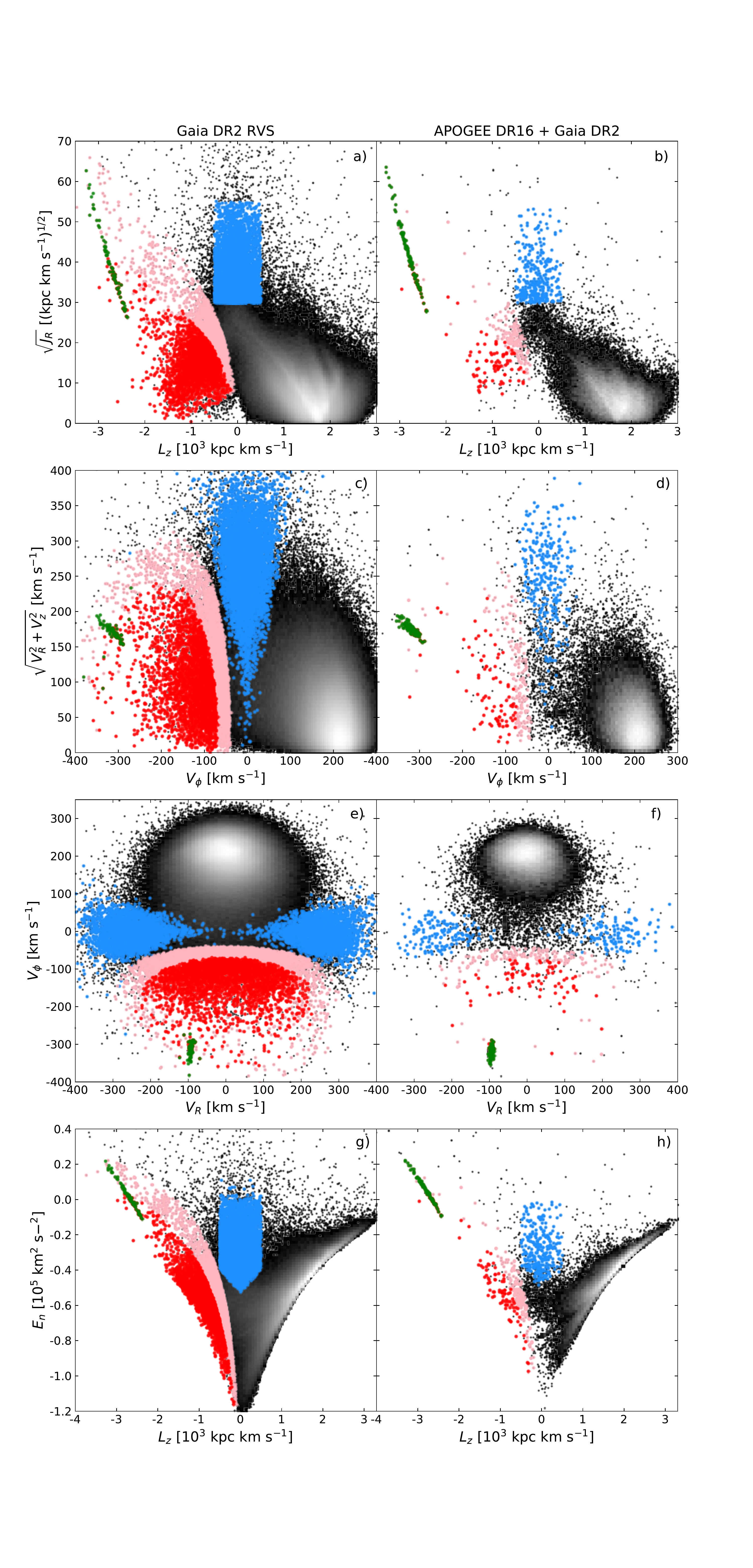}
\caption{The log(density) kinematic distributions (see \S \ref{sec:data}) of \Gaia DR2 RVS (left panels) and APOGEE DR16 + \Gaia DR2 (right panels). The selected GSE stars are shown in blue, see \S \ref{sec:sel_GSE}. Stars using the Sequoia 06 selection are shown in red, and Sequoia selection in red + pink, see \S \ref{sec:findseq} for details on the two Sequoia selections. NGC 3201 is easily identified in both datasets as the clump of stars with high $E_n$ and highly retrograde and are shown in green. }
\label{fig:kinematics}
\end{figure*}

\begin{figure*}
\includegraphics[clip,width=1.1\hsize,angle=0,trim=1cm 0cm 0cm 0cm] {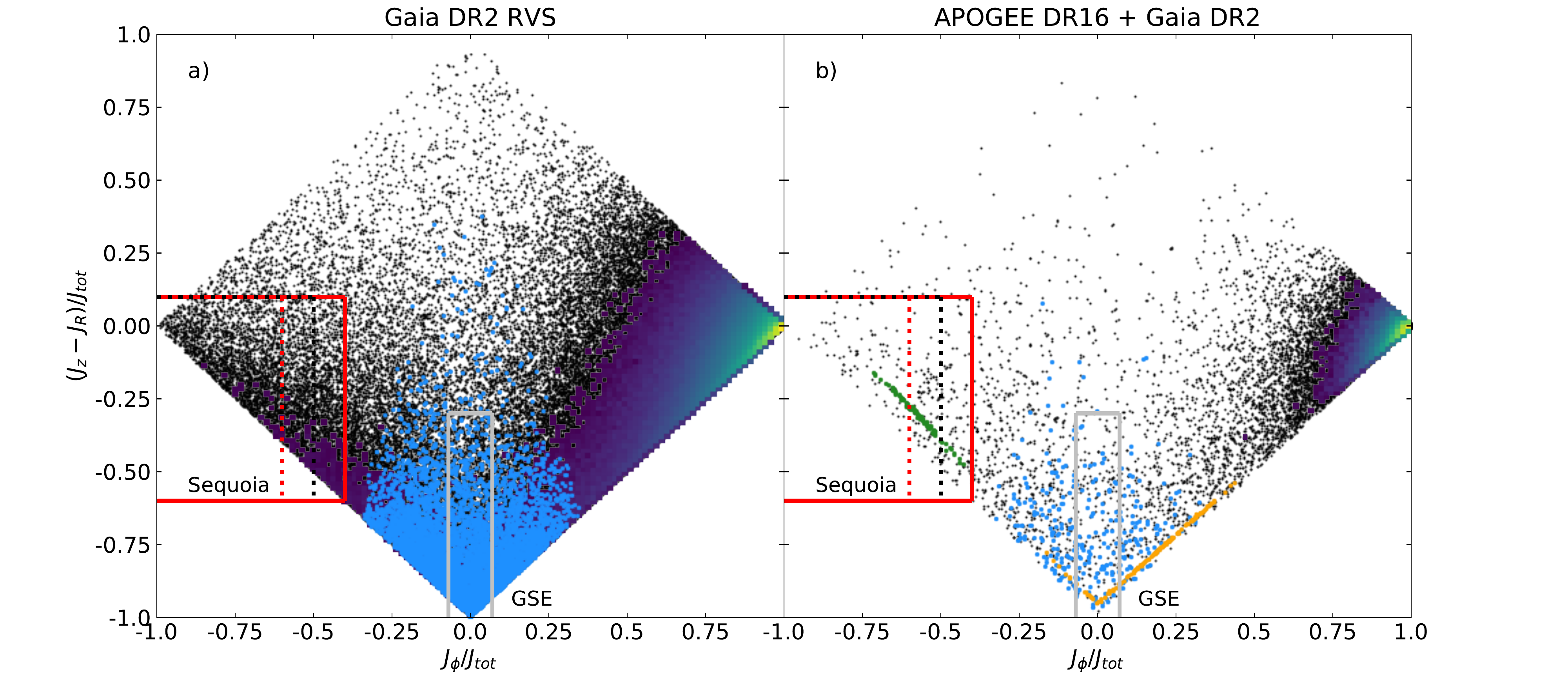}
\caption{The action space map of \Gaia DR2 RVS (a) and APOGEE DR16 + \Gaia DR2 (b) data. The Sequoia selection used in this work is shown in red, $J_{\phi}/J_{tot} = -0.4$ is solid and $J_{\phi}/J_{tot} < -0.6$ is a dotted line. The original \citet{Myeong2019} Sequoia selection is shown as the black dotted line. The \citet{Myeong2019} Sausage selection is shown as a grey line. The GSE stars identified in this paper selected using the adjusted \citet{Feuillet2020} criteria are shown as blue points. NCG 3201 cluster members are shown as green points and M4 cluster members are shown as orange points (b).}
\label{fig:square}
\end{figure*}

\section{Membership Criteria and Metallicity Distribution Functions}
\label{sec:select}

The possible GSE and Sequoia member stars are selected kinematically and some refinement in selection is done based on elemental abundances. Figure~\ref{fig:kinematics} shows the kinematic distributions of the full \Gaia DR2 RVS (left) and APOGEE DR16 + \Gaia DR2 (right) samples in four different spaces: a, b) $\sqrt{J_R}$ vs $L_z$, c, d) $V_{\phi}$ vs $V_R$, e, f) $\sqrt{V_R^2 + V_z^2}$ vs $V_{\phi}$, and g, h) $E_n$ vs $L_z$. 

The majority of stars have disc-like kinematics (motions dominated by rotational velocity in the prograde direction), but both samples extend to high $J_R$, high $E_n$, and highly retrograde orbits, regions where accreted populations are most apparent. The stars selected as GSE are shown in blue and the Sequoia candidate members identified in this work are shown as red and pink points, see Section \ref{sec:findseq}. Details of the selection criteria are described below and summarized in Table \ref{tab:selections}. In both the \Gaia DR2 RVS and APOGEE DR16 + \Gaia DR2 data, the globular cluster NGC~3201 is striking as the highly retrograde, high energy overdensity shown in green. This cluster has been associated with the Sequoia population by \citet{Myeong2019} using actions and energy.

\begin{table}
\caption{APOGEE Selection Criteria \label{tab:quality}}
\centering
\begin{tabular}{l l}
\hline \hline 
Property & Value \\
\hline
SNR & $> 80$  \\
\teff & $< 6000$  \\
\logg & $< 3.5$  \\
STARFLAG & $=0$  \\
ASPCAPFLAG & $=0$  \\
$\sigma_{\mbox{[Mg/Fe]}}$ & $< 0.05$ \\
\hline
\end{tabular}
   \begin{tablenotes}
    \item Note. --- Selection criteria for stars from the APOGEE DR16 + \Gaia DR2 sample, see Section \ref{sec:data}.
    \end{tablenotes}
\end{table}

\subsection{Identifying GSE Members}
\label{sec:sel_GSE}

The GSE stars were selected using angular momentum ($L_z$) and radial action ($\sqrt{J_R}$), Figure \ref{fig:kinematics} a and b, following the `clean' GSE selection from \citet{Feuillet2020}. Using SkyMapper [Fe/H] \citep{Casagrande2019} crossmatched with \Gaia DR2 RVS, \citet{Feuillet2020} found that stars with $30 \leq \sqrt{J_R} \leq 50$ (kpc km s$^{-1})^{1/2}$ and $-500 \leq L_z \leq 500$ kpc km s$^{-1}$ present the least-contaminated sample of kinematically selected GSE stars based on the [Fe/H] distribution. With the order of magnitude larger \Gaia DR2 RVS only sample, an extension of the GSE structure at higher $J_R$ is apparent. These stars were not present in the \citet{Feuillet2020} study because the SkyMapper sample was limited to parallax uncertainties less than $10\%$. Using the high $J_R$ APOGEE DR16 + \Gaia DR2 stars, we confirm that the [Fe/H] distribution of these stars is consistent with the main GSE population. 
We therefore extend our selection of GSE stars from both the \Gaia DR2 RVS and APOGEE DR16 + \Gaia DR2 datasets to $30 \leq \sqrt{J_R} \leq 55$ (kpc km s$^{-1})^{1/2}$.

\begin{figure*}
\includegraphics[clip,width=0.99\hsize,angle=0,trim=3cm 0cm 3.5cm 1cm] {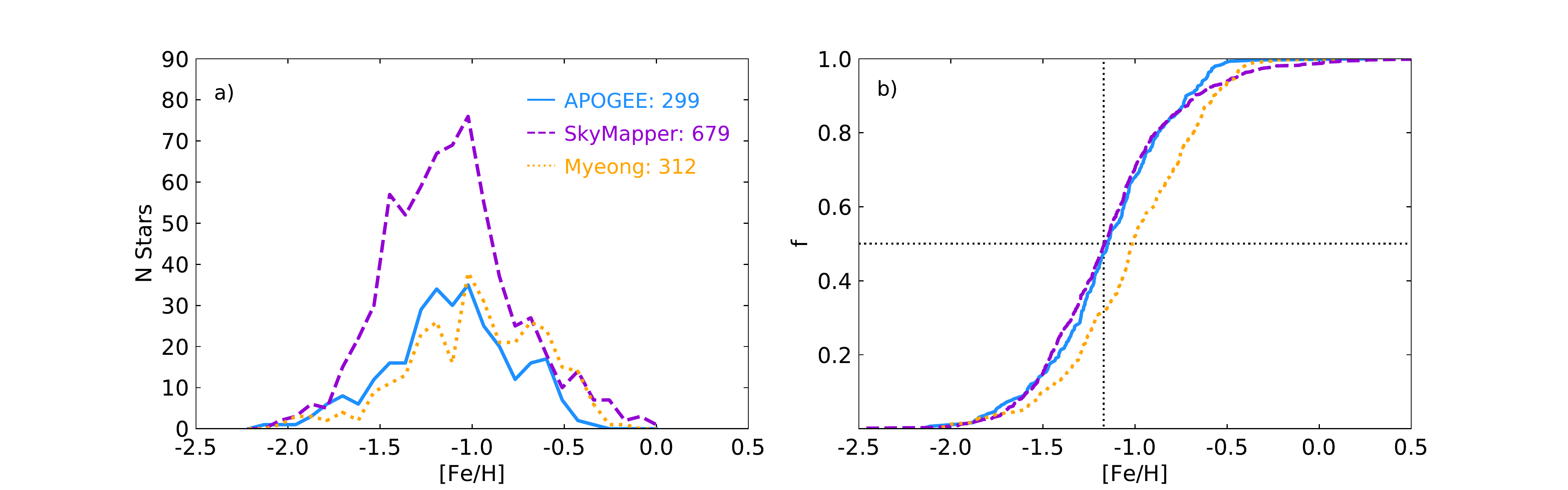}
\caption{The MDF (a) and CDF (b) of GSE using SkyMapper photometric metallicities (purple dashed, \citet{Feuillet2020}), APOGEE DR16 + \Gaia DR2 [Fe/H] (blue solid), and \citet{Myeong2019} GSE selection applied to APOGEE (yellow dotted, M19 GSE). In panel b, the black dotted lines indicate the 50\% cumulative fraction and the median metallicity of the SkyMapper data, [Fe/H] $= -1.17$. The number of stars in each sample is given in the top right corner of panel a.}
\label{fig:skyMDF}
\end{figure*}

\citet{Myeong2019} provide a GSE selection in the so-called action space map, which uses $J_{\phi}/J_{tot}$ on the horizontal axis and $(J_z - J_R)/J_{tot}$ on the vertical axis, see Figure \ref{fig:square}. We find that while GSE stars selected following the prescription in \citet{Myeong2019} do overlap in all kinematic spaces with the stars selected using the \citet{Feuillet2020} criteria, neither is a subset of the other. The \citet{Feuillet2020} selection is more extended in $L_z$, while \citet{Myeong2019} selects stars with lower $J_R$ and $E_n$. Figure \ref{fig:square} shows the GSE stars selected in the current work as blue points while the \citet{Myeong2019} `Sausage' selection is much narrower, as indicated by the grey lines. We find that GSE stars selected from the APOGEE DR16 + \Gaia DR2 sample using the \citet{Myeong2019} criteria have a mean [Fe/H] that is $0.15$ dex higher than when selected using \citet{Feuillet2020} criteria. This is apparent in Figure \ref{fig:skyMDF} b.

A second cluster streak can be seen along the bottom right side of panel b in Figure \ref{fig:square} overlapping with the GSE points, and is also identifiable in several panels of Figure \ref{fig:kinematics}. This is the globular cluster M4, which has a mean [Fe/H] of $-1.02$ based on APOGEE metallicity and has not been associated with GSE.

Our final GSE sample selected from the \Gaia DR2 RVS sample contains 5478 stars and our APOGEE DR16 + \Gaia DR2 sample of the GSE contains 323 stars before possible cluster stars are removed and 299 stars without stars in cluster fields. Kinematic selection criteria are given in Table \ref{tab:selections}.
We explore the abundance trends and ages of the GSE samples in Sections \ref{sec:abund} and \ref{sec:accrete}.

\begin{table*}
\caption{Kinematic Selection Criteria \label{tab:selections}}
\centering
\begin{tabular}{l l r l l l l}
\hline \hline 
Population & Source & & Selection &  & \multicolumn{2}{c}{N Stars} \\
 & & & & & APOGEE & RVS \\
\hline
GSE & This work & $-500 \leq$ & $L_z$ & $\leq 500$ & 299 & 5478 \\
& &  $30 \leq$ & $\sqrt{J_R}$ & $\leq 55$ & \\
 & \citet{Feuillet2020} & $-500 \leq$ & $L_z$ & $\leq 500$ & 292 & 5271 \\
& &  $30 \leq$ & $\sqrt{J_R}$ & $\leq 50$ & \\
 & \citet{Myeong2019} & $-0.07 <$ & $J_{\phi}/J_{\mbox{tot}}$ & $< 0.07$ & 312 & 4994 \\
& & $-1.0 <$ & $(J_z - J_R)/J_{\mbox{tot}}$ & $< -0.3$ & \\
Sequoia & This work & $-1.0 <$ & $J_{\phi}/J_{\mbox{tot}}$ & $< -0.4$ & 244 & 6379 \\
& & $-1.0 <$ & $(J_z - J_R)/J_{\mbox{tot}}$ & $< 0.1$ & \\
 & \citet{Myeong2019} & $-1.0 <$ & $J_{\phi}/J_{\mbox{tot}}$ & $< -0.5$ & 155 & 4293 \\
& & $-1.0 <$ & $(J_z - J_R)/J_{\mbox{tot}}$ & $< 0.1$ & \\
Sequoia 06 & This work & $-1.0 <$ & $J_{\phi}/J_{\mbox{tot}}$ & $< -0.6$ & 79 & 2790 \\
& & $-1.0 <$ & $(J_z - J_R)/J_{\mbox{tot}}$ & $< 0.1$ & \\

\hline
\end{tabular}
\vspace{-0cm}
\begin{tablenotes}
\item Note. --- Numbers of stars given for the APOGEE DR16 + \Gaia DR2 samples have stars in certain fields removed, as listed in Table \ref{tab:fields}.
\end{tablenotes}
\end{table*}

\subsection{Finding the Sequoia}
\label{sec:findseq}

To select Sequoia stars, we follow \citet{Myeong2019} and use the action space map. In this space, \citet{Myeong2019} select Sequoia stars as $J_{\phi}/J_{tot} < -0.5$ and $(J_z - J_R)/J_{tot} < 0.1$ (black dotted lines in Figure \ref{fig:square}). In Figure \ref{fig:square}, NGC~3201 is easily identifiable in the Sequoia selection region of the APOGEE DR16 + \Gaia DR2 data as an overdensity streak. Cluster members, identified in APOGEE DR16 + \Gaia DR2 using [Fe/H] and kinematics, are shown as green points and extend beyond the \citet{Myeong2019} selection region. We therefore test different Sequoia selection limits in $J_{\phi}/J_{tot}$. Our main Sequoia selection extends to $J_{\phi}/J_{tot} < -0.4$ in order to include all NGC~3201 member stars. This selection is indicated by the solid red lines in Figure~\ref{fig:square} and plotted as the red + pink points in Figure \ref{fig:kinematics}. An alternative, more conservative Sequoia selection, which is limited to $J_{\phi}/J_{tot} < -0.6$, is indicated by the dotted red lines in Figure~\ref{fig:square} and the red points in Figure \ref{fig:kinematics}. We refer to the sample selected in this way as Sequoia 06, see Table \ref{tab:selections}. We evaluate these selection limits using the MDFs and abundance distributions, see discussion in Section \ref{sec:abund}.

The final Sequoia and Sequoia 06 samples selected from the \Gaia DR2 RVS catalogue contain 6379 and 2790 stars, respectively. The APOGEE DR16 + \Gaia DR2 samples of the Sequoia and Sequoia 06 contain 372 and 116 stars, respectively, before possible cluster stars are removed and 244 and 79 stars, respectively, without stars in cluster fields. We explore the abundance trends and ages of the Sequoia and Sequoia 06 samples in Sections \ref{sec:abund} and \ref{sec:accrete}.



\section{Characterizing Gaia-Sausage-Enceladus and Sequoia} 
\label{sec:abund}

\subsection{Metallicity Distribution Functions}

\subsubsection{GSE}

Figure \ref{fig:skyMDF} compares the MDF of the GSE stars selected from APOGEE DR16 + \Gaia DR2 (blue) with the MDF found by \citet{Feuillet2020} using SkyMapper photometric metallicities (purple). In Figure \ref{fig:skyMDF} b, which shows the cumulative MDF (CDF), the dotted lines indicate the 50\% cumulative fraction and the median [Fe/H] of the SkyMapper sample, $-1.17$ dex. The APOGEE DR16 + \Gaia DR2 median [Fe/H] is almost exactly the same as the one found using SkyMapper data. The MDF shapes are also very similar, the main difference being that the SkyMapper MDF has a metal-rich tail while the APOGEE DR16 + \Gaia DR2 MDF has a metal-rich secondary peak around [Fe/H] $\sim -0.6$. The \citet{Myeong2019} GSE selection applied to APOGEE (yellow) also has a metal-rich peak, however, both of these metal-rich peaks are within the Poisson noise of their distributions. The agreement serves as a good validation of both the SkyMapper metallicities and the GSE selection criteria of \citet{Feuillet2020}. 

\begin{figure*}
\includegraphics[clip,width=0.99\hsize,angle=0,trim=2cm 0.5cm 3cm 2cm] {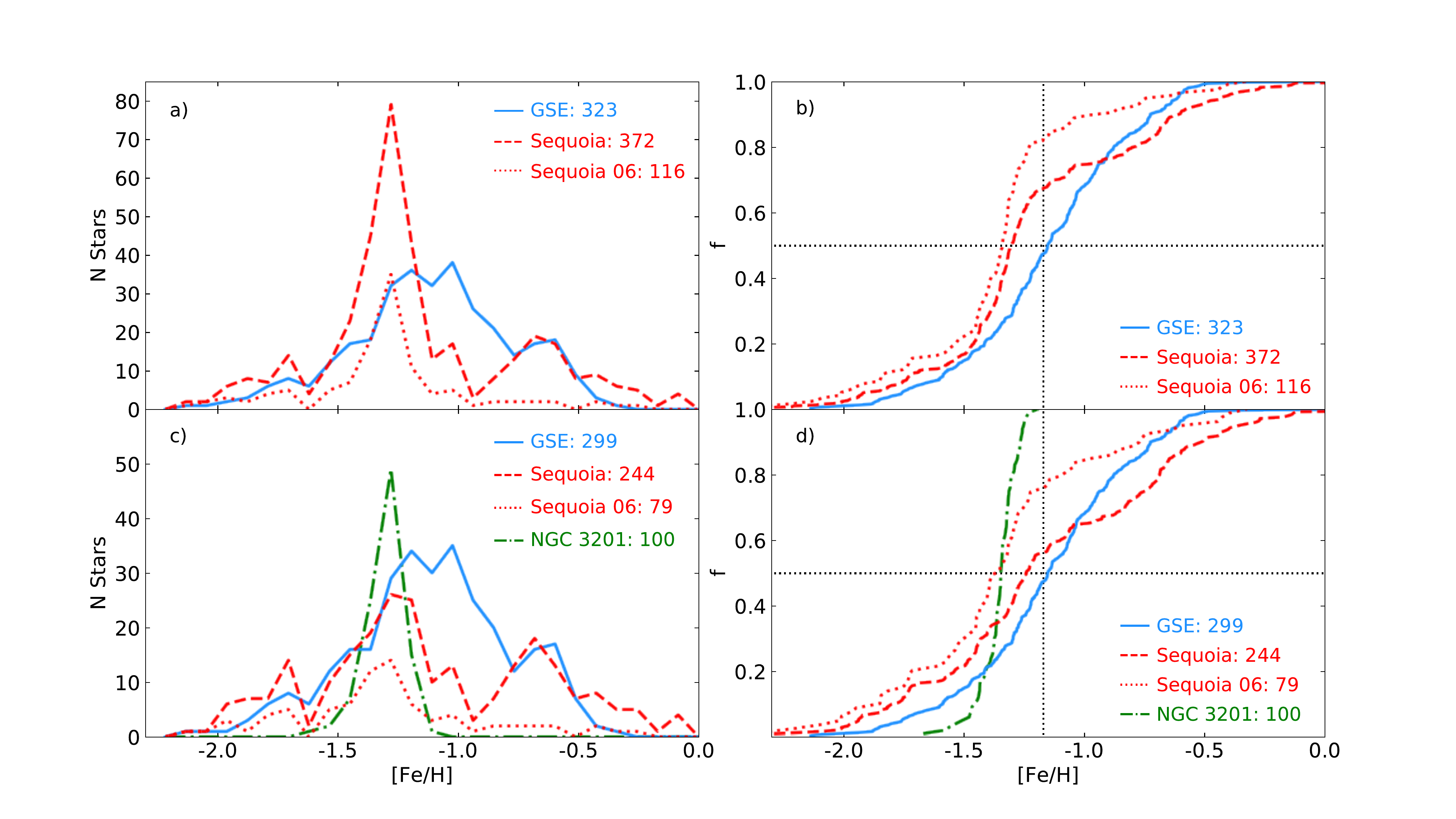}
\caption{The MDFs (a, c) and CDFs (b, d) of GSE (blue solid), Sequoia (red dashed), and Sequoia 06 (red dotted) using APOGEE DR16 + \Gaia DR2 [Fe/H]. The top panels (a, b) show the metallicity distributions before stars were removed based on the APOGEE `FIELD' parameter, which is used in Section \ref{sec:ages}. The bottom panels (c, d) show the APOGEE DR16 + \Gaia DR2 sample with stars in select fields removed, see Section \ref{sec:data} and Table \ref{tab:fields}. In the bottom panels NGC 3201 cluster members are shown in a green dot dashed line. The horizontal and vertical dotted lines indicate the 50th percentile and [Fe/H] $= -1.17$, respectively. [Fe/H] $= -1.17$ is the mean metallicity found for GSE using the SkyMapper data by \citet{Feuillet2020}.}
\label{fig:MDF}
\end{figure*}

Figure \ref{fig:MDF} shows the MDF and CDF of GSE stars in blue using APOGEE DR16 + \Gaia DR2 [Fe/H]. The top panels (a, b) show the kinematically selected APOGEE DR16 + \Gaia DR2 GSE sample with only the quality selection criteria imposed (Table \ref{tab:quality}), but no stars removed based on the APOGEE `FIELD' parameter (Table \ref{tab:fields}). This distribution is used when determining the GSE population age from the \Gaia DR2 RVS sample, see Section \ref{sec:ages}. The bottom panels (c, d) show the APOGEE DR16 + \Gaia DR2 GSE sample with stars in select fields removed, as described in Section \ref{sec:data}. We find that GSE has a mean [Fe/H] of $-1.15$ dex in both cases. The metal-rich peak is also present in Figure \ref{fig:MDF} a and c although 24 stars have been removed in Figure \ref{fig:MDF} c.

\subsubsection{Sequoia}

Figure \ref{fig:MDF} shows the MDF and CDF of Sequoia stars in red. Figure \ref{fig:MDF} a and b show the kinematically selected APOGEE DR16 + \Gaia DR2 Sequoia and Sequoia 06 samples with only the quality selection criteria imposed (Table \ref{tab:quality}), but no stars removed based on the APOGEE `FIELD' parameter (Table \ref{tab:fields}). Figure \ref{fig:MDF} c and d show the APOGEE DR16 + \Gaia DR2 Sequoia and Sequoia 06 samples with stars in select fields removed, as described in Section \ref{sec:data}. NGC 3201 cluster members are shown separately in green. The alternative Sequoia selection, Sequoia 06, is shown by the dotted red lines. We find that Sequoia has a mean/median [Fe/H] of $-1.2/-1.3$ dex when NGC 3201 is included and $-1.15/-1.24$ dex without it. In both cases, a metal-rich secondary peak is present around [Fe/H] $\sim -0.6$ and a tertiary metal-poor peak around [Fe/H] $\sim -1.7$, however, this metal-poor peak is within the Poisson noise. When NCG 3201 is not included in the main Sequoia sample, the main peak is consistent with the very narrow peak of NGC 3201, which has a mean [Fe/H] of $-1.35$ dex and a standard deviation of $0.08$ dex.

The Sequoia 06 selection has a mean [Fe/H] of $-1.35$ dex both with and without NGC 3201. The main peak of the Sequoia and Sequoia 06 samples are consistent. There is no peak at $-0.6$ dex in the Sequoia 06 sample, although metal-rich starts are present in small numbers, but the $-1.7$ dex peak is present. In both the Sequoia and Sequoia 06 samples, the metal-rich contribution is particularly striking in the CDFs after the sharp increase around [Fe/H] $\sim-1.3$.

\subsubsection{Comparisons}
\label{sec:comp}

Using APOGEE DR14 data, \citet{Myeong2019} found that GSE and Sequoia have different [Fe/H] distributions.
We visit this finding with a larger sample from APOGEE DR16 + \Gaia DR2. For our samples, the main peak of the GSE MDF is broader and more metal-rich than the main peak of the Sequoia MDF. However, the range of metallicities covered by the Sequoia sample is slightly larger than the GSE sample, extending to both higher and lower [Fe/H].

Both have a secondary metal-rich peak around [Fe/H] $\sim -0.6$, which is not present in the more restricted Sequoia 06 sample. The similarity of this peak between GSE and Sequoia suggests the stars comprising the peak may be {\it in situ} Milky Way stars such as kinematically heated thick disc stars. In addition, many of the stars contributing to this peak in both GSE and Sequoia have disc-like elemental abundances, see Figure \ref{fig:MgFe} and Section \ref{sec:accrete}. This supports Sequoia 06 as the cleaner sample of Sequoia stars as the metal-rich (likely {\it in situ}) peak is not present.



\subsection{Elemental Abundances}

Distinct accreted populations can often be identified by their elemental abundance patterns. Many studies have shown that the elemental abundance patterns of accreted populations in the Milky Way are different from the main {\it in situ} population's elemental abundances (e.g. \citealt*{Mcwilliam2013}; \citealt{Hasselquist2017}; \citealt{Das2020}; \citealt{Hayes2018}) and local dwarf galaxies are known to have distinct elemental abundance patterns \citep[see][]{Reichert2020}. This is thought to be caused primarily by different star formation rates, however, there could be other contributing factors such as variations in the initial mass function \citep[e.g.][]{Carlin2018}. 
In addition to having different peak [Fe/H], \citet{Myeong2019} also conclude that GSE and Sequoia have distinct elemental abundance patterns in [Mg/Fe] and [Al/Fe]. 

\begin{figure*}
\includegraphics[clip,width=0.49\hsize,angle=0,trim=0.2cm 2.5cm 0.5cm 3cm] {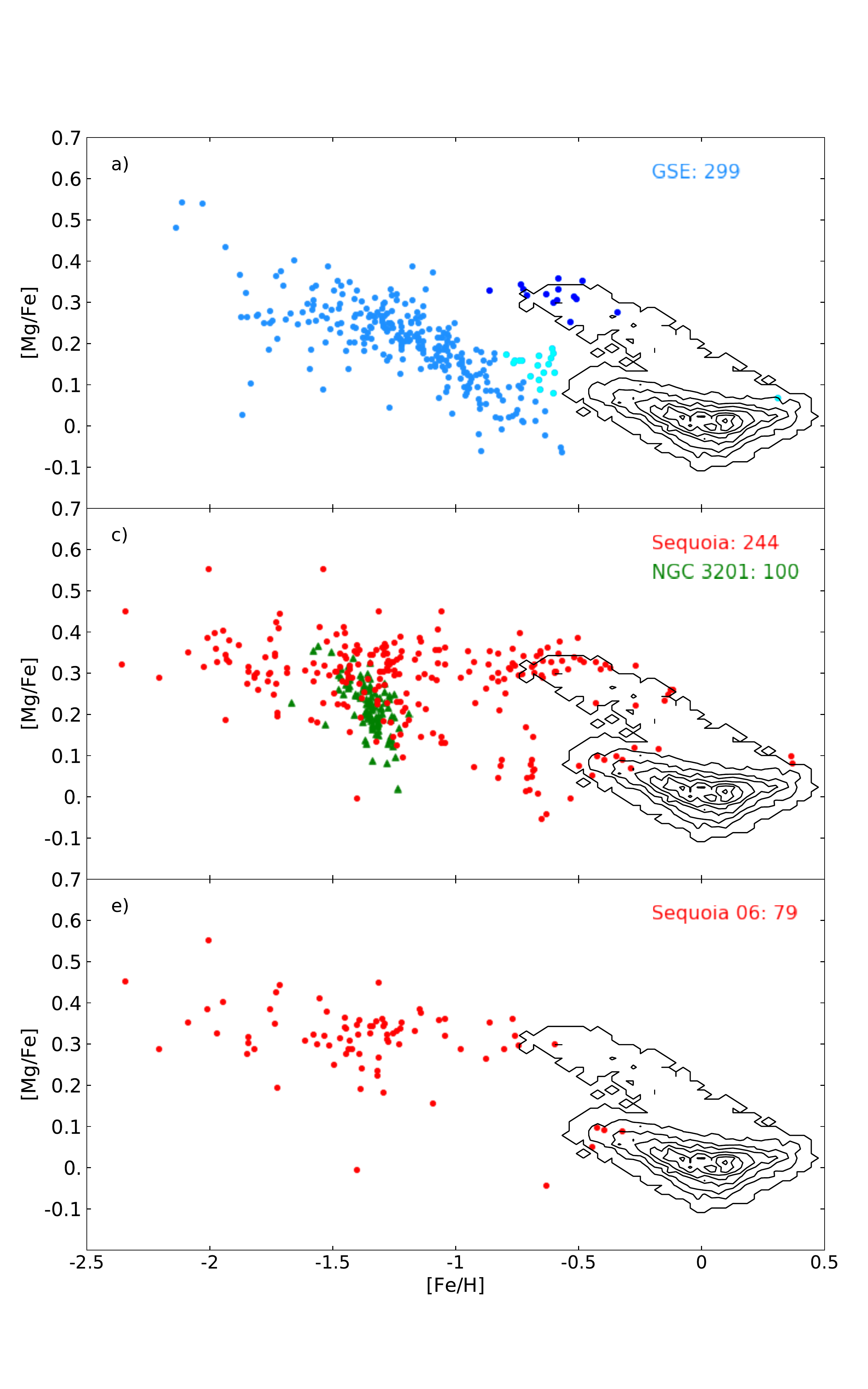}
\includegraphics[clip,width=0.49\hsize,angle=0,trim=0.2cm 2.5cm 0.5cm 3cm] {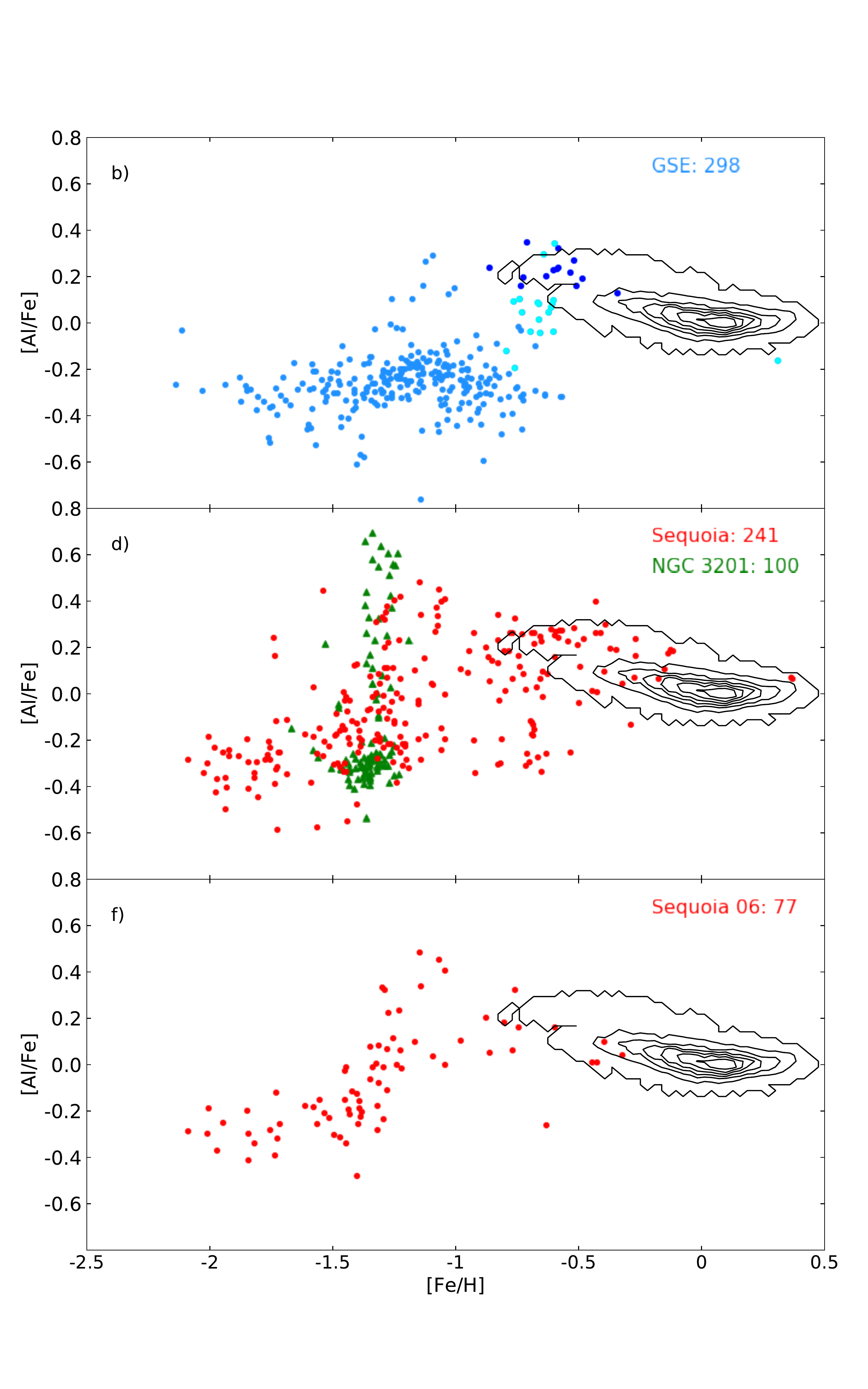}
\caption{APOGEE DR16 + \Gaia DR2 abundance distributions of GSE (blue), Sequoia (red), and NGC 3201 members (green). The Sequoia 06 selection is shown in the bottom panels. [Mg/Fe] vs [Fe/H] is shown in the left panels and [Al/Fe] vs [Fe/H] in the right panels. The black contours indicate the solar neighborhood distribution. The number of stars shown for each sample is indicated. In panels a and b, stars possibly belonging to the high- or low-[Mg/Fe] sequences of the MW disc are indicated in dark and light blue, respectively.}
\label{fig:MgFe}
\end{figure*}

Figure \ref{fig:MgFe} shows [Mg/Fe] and [Al/Fe] abundance distributions as a function of [Fe/H] for GSE (blue), Sequoia (red), and NGC 3201 (green) using APOGEE DR16 + \Gaia DR2. For reference, the black contours show the distribution of solar neighborhood ($7 \leq$ \Rgal $\leq 9$, \z $\leq 0.5$) APOGEE DR16 + \Gaia DR2 giant stars meeting with same quality criteria as the GSE and Sequoia samples, see Section \ref{sec:data}.

\subsubsection{GSE}

The GSE population lies mainly along a single narrow track in [Mg/Fe] vs [Fe/H] (Figure \ref{fig:MgFe} a), with a `knee' at [Fe/H] $\sim -1.4$ and a metal-poor plateau at [Mg/Fe] $\sim -0.25$. Some stars are present at high [Mg/Fe] and high [Fe/H], highlighted in darker blue, and are consistent with the high-[Mg/Fe] disc distribution. Other possible {\it in situ} stars are highlighted in lighter blue and are consistent with the metal-poor end of the low-[Mg/Fe] disc.
Together, these disc-like stars contribute to the $-0.6$ dex peak in the GSE MDF that was suggested to comprise {\it in situ} stars, see Section \ref{sec:comp}. Both subsamples are also indicated in panel b and have high [Al/Fe], consistent with the Milky Way disc populations.

The [Al/Fe] of most of the GSE stars is much lower than the Milky Way disc, at [Al/Fe] $\sim -0.3$, consistent with GSE in \citet{Myeong2019} and that of other known accreted populations \citep[e.g. Sagittarius,][]{Hasselquist2017}. The distribution is approximately flat in [Al/Fe] vs [Fe/H] (Figure~\ref{fig:MgFe} b) with a slight peak in [Al/Fe] around [Fe/H]~$\sim-1.2$. Some stars have higher [Al/Fe] abundance ratios. In particular, some stars at [Fe/H] $\sim -0.6$ have an [Al/Fe] consistent with the Milky Way disc distributions. As indicated in Figure \ref{fig:MgFe}, these stars also have [Mg/Fe] values consistent with the MW disc populations. The difference in abundance at the metal-rich end is larger in [Al/Fe] than in [Mg/Fe], particularly for the low-[Mg/Fe] stars, suggesting that Al may be a better diagnostic of accreted origins in individual stars.

\subsubsection{Sequoia}

Figure \ref{fig:MgFe} c and d show the main Sequoia selection while Figure \ref{fig:MgFe} e and f show the Sequoia 06 selection. The Sequoia stars appear to follow multiple tracks in [Mg/Fe] vs [Fe/H], one high [Mg/Fe] track consistent with the canonical halo/thick disc track and a less populated lower [Mg/Fe] track similar to the track followed by GSE. The dual tracks suggest there are multiple populations contributing to the Sequoia sample selected here. 
NGC 3201 (shown in green triangles) sits at the metal-poor end of the Sequoia lower [Mg/Fe] sequence, slightly below the majority of the Sequoia stars in [Mg/Fe].

The Sequoia 06 distribution lies mainly along the high-Mg track, but still has a few stars along the low-Mg track. Comparing the abundance distributions of these two Sequoia selections, we find that the more conservative Sequoia 06 selection produces a distribution that more closely resembles a single population, especially in comparison with the low apparent contamination of the GSE distribution. The high-Mg track resembles the elemental abundance pattern seen generally in the Milky halo and the [Mg/Fe] trend predicted by chemical evolution models of the Milky Way \citep[e.g.][]{Romano2010, Andrews2017}. However, we note the large variations in abundance trend predictions of chemical evolution models resulting in small changes to the model assumptions such as star formation rate or theoretical yields \citep[see][for a review]{Matteucci2021}. While chemical evolution models have been made to fit observations of dwarf galaxies and GSE \citep[e.g.][]{Vincenzo2019}, these efforts are intended to constrain the properties of the progenitor galaxy, not provide predictions of the elemental abundance trends of low-mass systems.

\begin{figure*}
\includegraphics[clip,width=0.99\hsize,angle=0,trim=0cm 0cm 0cm 0cm] {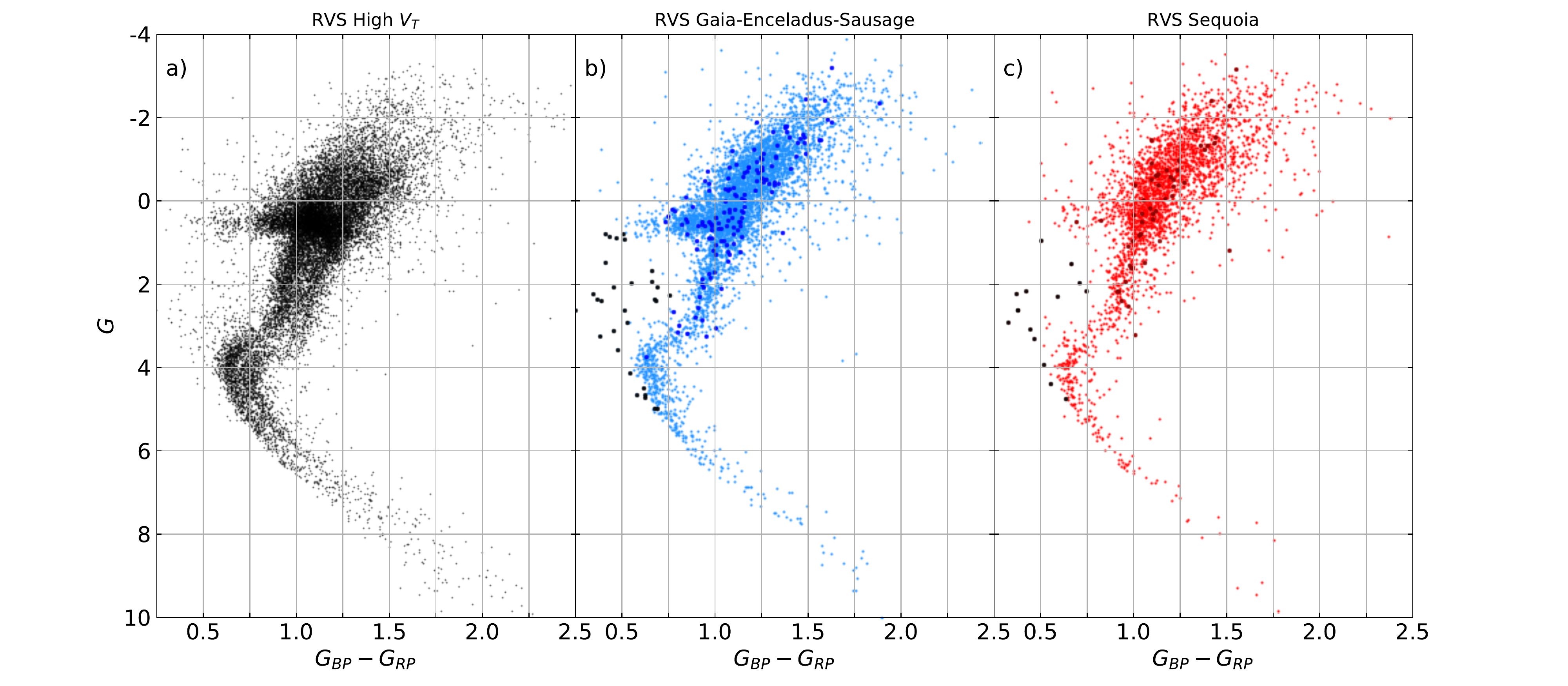}
\caption{Dereddened CMDs of the \Gaia DR2 RVS high $V_T$ stars, \Gaia DR2 RVS GSE, and \Gaia DR2 RVS Sequoia 06. GSE and Sequoia stars contained in the APOGEE samples are indicated in dark blue and dark red, respectively. GSE and Sequoia stars with individual ages less than 6 Gyr are indicated in black.}
\label{fig:CMD}
\end{figure*}

Similarly to the [Mg/Fe] distribution, the Sequoia stars have an [Al/Fe] vs [Fe/H] distribution suggestive of multiple populations contributing, a low-[Al/Fe] population and a high-[Al/Fe] population. The low [Al/Fe] sequence extends to lower [Fe/H], while the high [Al/Fe] stars are consistent with the Milky Way disc populations and extend to higher [Fe/H]. The Sequoia 06 selection does not have the low [Al/Fe] extension at higher [Fe/H] and also has few stars overlapping with the main disc distribution. 
NCG 3201 has an extended [Al/Fe] distribution ($\sim1.0$ dex), consistent with known abundance patterns in globular clusters \citep[e.g.][]{Meszaros2020}. Most of the stars, considered to be the first generation stars, have Al abundances lower than the Sequoia stars at the same [Fe/H]. 

From the MDFs as well as the [Mg/Fe] and [Al/Fe] abundance distributions of the two Sequoia selections we examined, we conclude that the Sequoia 06 selection, although much smaller, provides a sample that has significantly less contamination. As a clean sample is crucial to determining the true Sequoia abundance patterns, we continue our analysis using the Sequoia 06 selection. 
The stars with [Fe/H]~$\sim-1$ to $-0.5$ with low [Mg/Fe] and low [Al/Fe] that are removed using the Sequoia 06 selection have abundance patterns consistent with GSE. These stars could be a retrograde extension of GSE, suggesting that the GSE stars occupy a large range of kinematic space. It is also conceivable that an elemental abundance gradient within Sequoia at the time of infall could result in a trend in [Al/Fe] with $J_{\phi}/J_{tot}$, causing the change in abundance distribution with a different kinematic selection.

As the Sequoia is usually characterized as a high energy population, we explore possible differences in the orbital energy distribution of the two [Mg/Fe] branches seen in the Sequoia distribution in Appendix \ref{ap:seq_en}. We find that the low [Mg/Fe] branch has an excess of high energy stars compared to the high [Mg/Fe] branch. However, we do not find that a Sequoia sample selected using orbital energy results in elemental abundances that are obviously better than the Sequoia 06 selection. We therefore continue our analysis using Sequoia 06 and present the high energy Sequoia distributions in Figures \ref{fig:seq_en} and \ref{fig:seq_en_abund}.

\begin{figure*}
\includegraphics[clip,width=0.4\hsize,angle=0,trim=0.5cm 1cm 1cm 1cm] {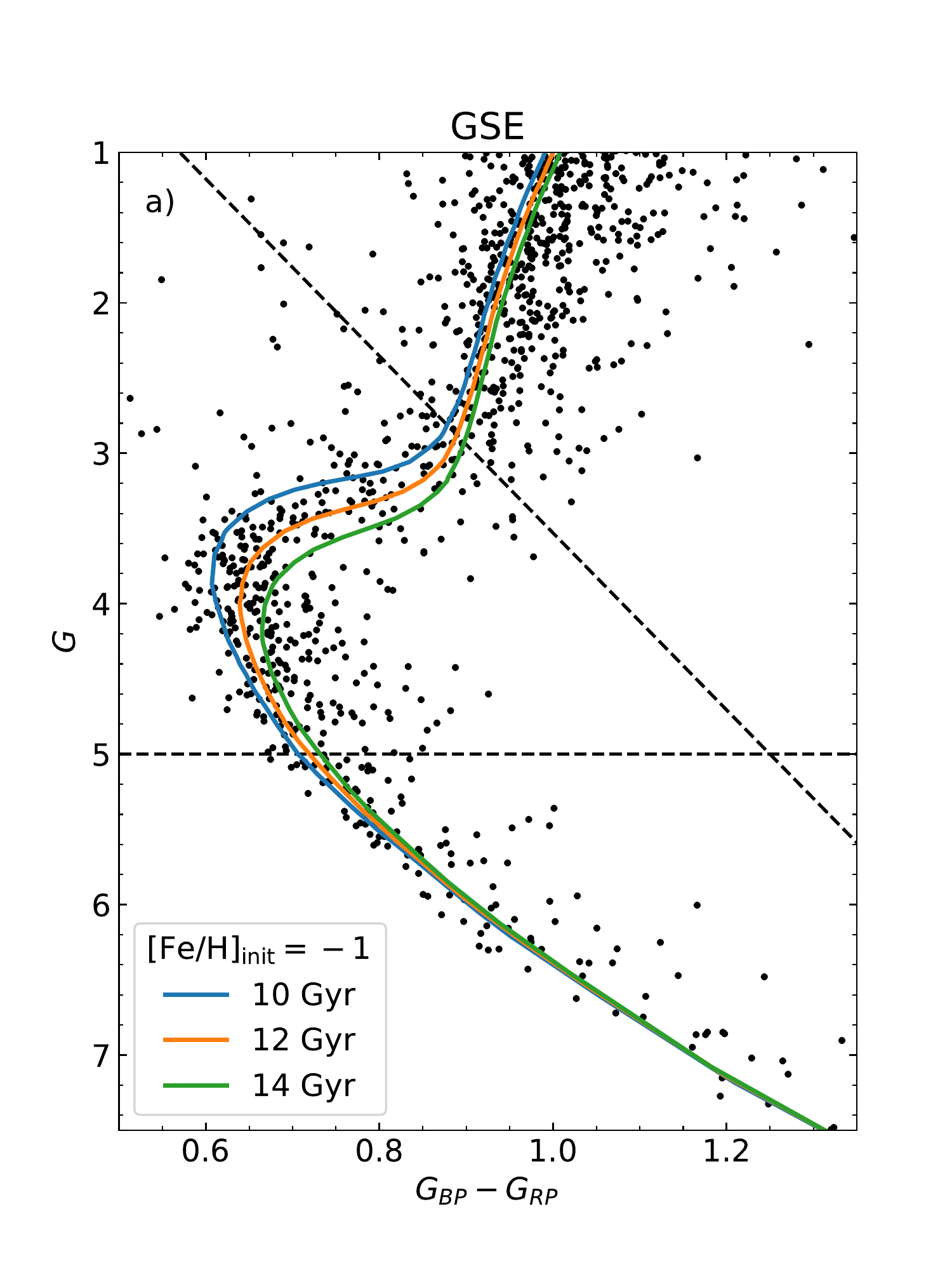}
\includegraphics[clip,width=0.4\hsize,angle=0,trim=0.5cm 1cm 1cm 1cm] {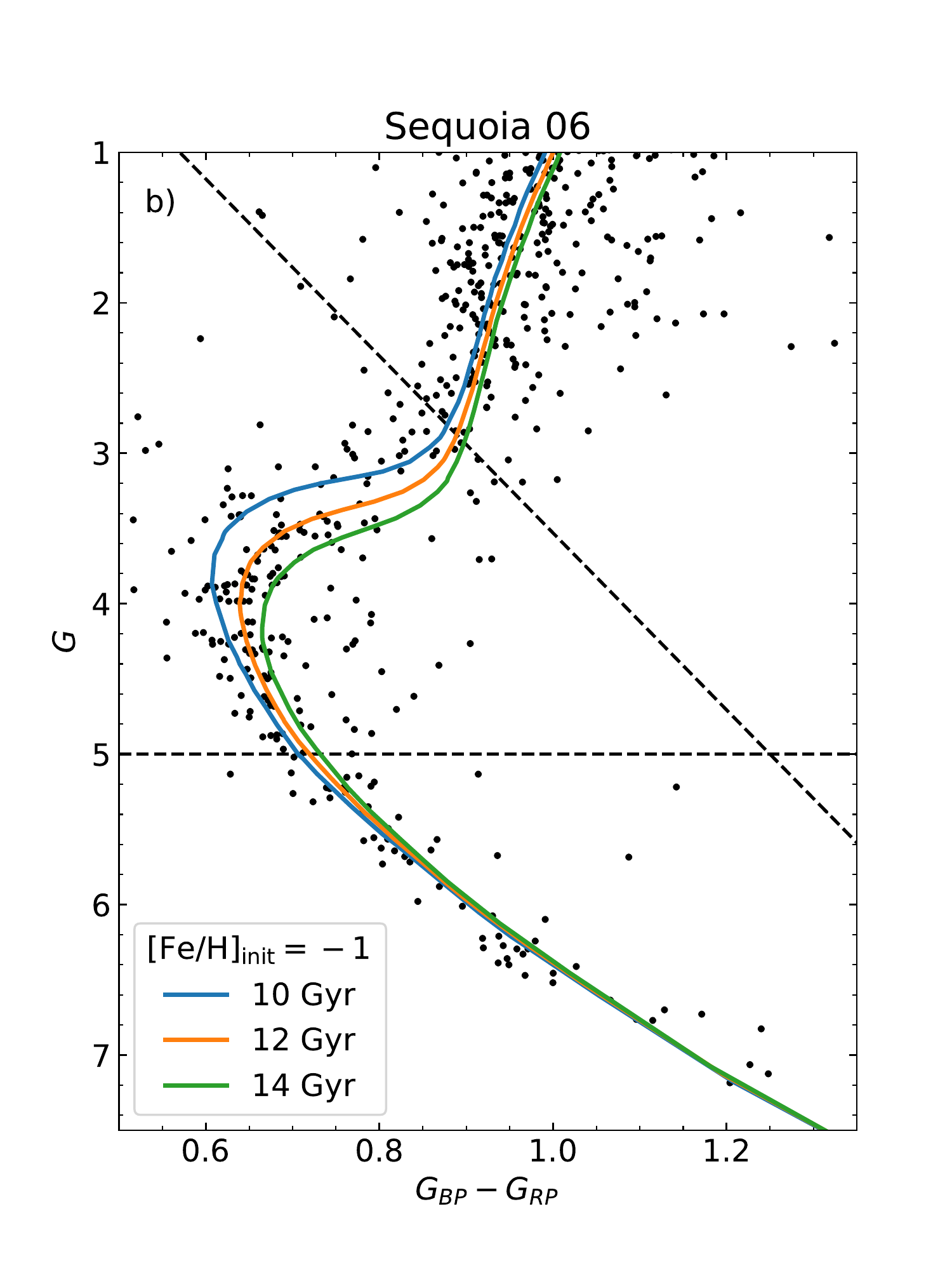}
\caption{The dereddened CMD of \Gaia DR2 RVS subgiant stars in the GSE (a) and Sequoia 06 (b) with 10, 12, and 14 Gyr PARSEC isochrones shown for [Fe/H]$_{\mbox{init}}$ = -1.0. 
The dashed lines show the selection of subgiant stars used in the age analysis.
}
\label{fig:iso_fit}
\end{figure*}

\subsection{Ages of GSE and Sequoia}
\subsubsection{CMD Inspection}

The samples selected from the \Gaia DR2 RVS catalog are large and contain full population sequences, including subgiant stars, which can be seen in the color-magnitude diagrams (CMDs) in Figure \ref{fig:CMD}. For the Sequoia population we use the Sequoia 06 selection, which Figure \ref{fig:MgFe} suggests contains less contamination from {\it in situ} Milky Way stars. 
Figure \ref{fig:CMD} a shows the high tangential velocity ($V_T > 200$ km s$^{-1}$) stars in \Gaia DR2 RVS, corresponding to the selection by \citet{Gaia2018b} that demonstrated the dual nature of such a sample. We find that while the Sequoia 06 sequence is slightly redder than the GSE sequence, both are consistent with the blue sequence of the high $V_T$ sample, which has been found to be more metal poor than the red sequence and likely mainly composed of accreted stars \citep[e.g.][]{Haywood2018, Sahlholdt2019, Gallart2019}.

The CMDs based on the \Gaia DR2 RVS data allow for age estimates using isochrone fitting.
We correct the Gaia magnitudes for extinction using reddening estimates, $E(B-V)$, from the Stilism project \citep{Lallement2019} and the mean extinction coefficients for the \Gaia passbands given by \citet{Casagrande2018}. Initial population age information can be gained from Figure \ref{fig:iso_fit}, which focuses on the subgiant region of GSE and Sequoia 06. PARSEC isochrones \citep{Bressan2012} of initial [Fe/H] $= -1$ are shown for $10$, $12$, and $14$ Gyr. The majority of stars in the turn off and subgiant region of GSE are consistent with an age of $10-14$ Gyr. The majority of Sequoia 06 stars in the turn off and subgiant region are consistent with an age of $12-14$ Gyr.

\subsubsection{Age Calculations}
\label{sec:ages}


For a more robust age analysis, PARSEC isochrones are fitted to the $G$-band magnitude, $G_{BP}-G_{RP}$ color, and parallax in order to obtain two-dimensional (age-metallicity) probability density functions (PDFs) for all stars following \citet{Howes2019}. For this analysis we assume a Salpeter initial mass function \citep{Salpeter1955} as the prior on mass.
This is done for both the GSE and Sequoia 06 \Gaia DR2 RVS samples separately selecting only stars around the turnoff of the CMD (indicated by the dashed lines in Figure \ref{fig:iso_fit}).


Since we do not have metallicity information for the full \Gaia DR2 RVS sample, we make the assumption that the metallicity distributions for the \Gaia DR2 RVS GSE and Sequoia 06 samples are the same as those of the corresponding APOGEE DR16 + \Gaia DR2 samples for the same kinematic selection. In the \Gaia DR2 RVS sample, no attempt has been made to remove stars belonging to clusters as was done with the APOGEE DR16 + \Gaia DR2 sample, described in Section \ref{sec:data}.
The [Fe/H] distributions assumed are the ones shown in panel a of Figure \ref{fig:MDF}. 
To include the assumed metallicity, we marginalize the metallicity in each individual age-metallicity PDF with a metallicity distribution prior as shown below. 
\begin{equation}
\mathcal{G}(\tau) = \int \mathcal{G}(\tau,[\textup{M}/\textup{H}])p([\textup{M}/\textup{H}])\textup{d}[\textup{M}/\textup{H}]
\end{equation}
where $\mathcal{G}(\tau, \mbox{[M/H]})$ is the 2D PDF from isochrone fitting and $p$([M/H]) is the metallicity prior.

The metallicity used in this prior is adjusted to account for the $\alpha$-enhancement of the GSE and Sequoia populations according to \citet*{Salaris1993},
\begin{equation}
\mbox{[M/H]} = \mbox{[Fe/H]} + \log(0.638 \times10^{\mbox{[Mg/Fe]}}+0.362)
\end{equation}
where [Mg/Fe] from APOGEE is used as the $\alpha$-abundance. This gives for each star an age PDF from which individual ages can be estimated, in this case we use the mode of the distribution. The uncertainties in the individual age estimates range from $10 - 50\%$, with a mode of $20\%$.

\begin{figure*}
\includegraphics[clip,width=0.49\hsize,angle=0,trim=0cm 0cm 0cm 0cm] {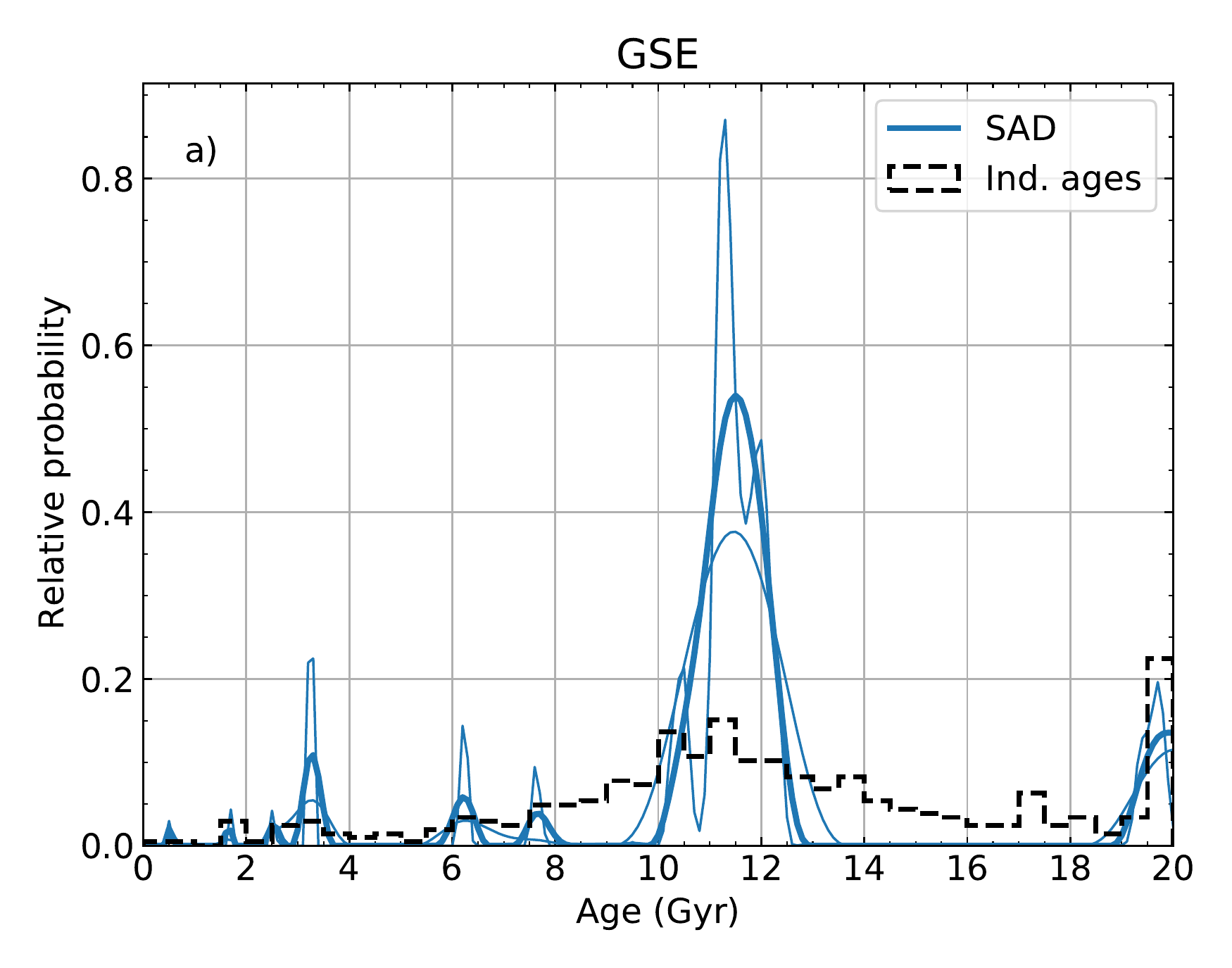}
\includegraphics[clip,width=0.49\hsize,angle=0,trim=0cm 0cm 0cm 0cm] {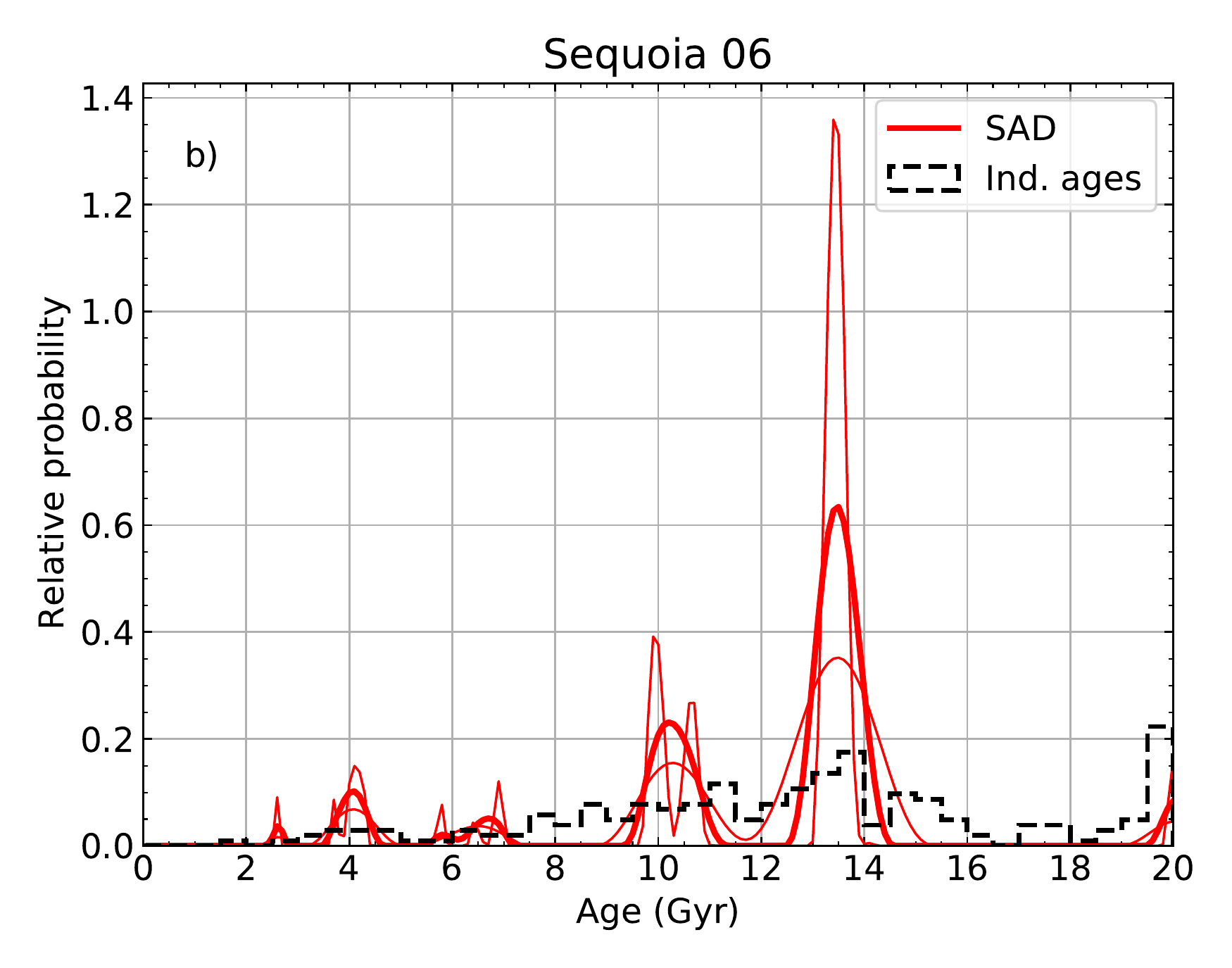}
\caption{Individual age distributions (dashed) and population SADs (solid) for \Gaia DR2 RVS GSE stars (a) and \Gaia DR2 RVS Sequoia 06 stars (b) assuming an MDF informed by the corresponding APOGEE DR16 + \Gaia DR2 sample. For the SADs, three values of the smoothing parameter ($\alpha$) are shown. The fiducial value $\alpha = 2000$ is shown by the thick line and the thin lines show $\alpha=0$ (no smoothing) and $\alpha=10^5$.}
\label{fig:ages}
\end{figure*}

In addition to the individual ages, we estimate the sample age distribution (SAD) for each of the GSE and Sequoia 06 samples, using the method of \citet{Sahlholdt2021} to combine the age PDFs. 
The SAD, $\phi(\tau)$, is calculated by minimising the negative log-likelihood given by
\begin{equation}
-\ln L = - \sum_{i} \ln\left( \int \mathcal{G}_{i}(\tau)\phi(\tau)\mathrm{d}\tau \right) + \alpha \int\left(s_{\tau}^2\dfrac{\partial^2\phi}{\partial\tau^2}  \right)^2 \mathrm{d}\tau \, ,
\end{equation}
where the second term determines the smoothness of the solution.
The smoothing is regulated in strength by the parameter $\alpha$ and $s_{\tau}=0.1$~Gyr is the resolution in age.
Both the individual age distribution and the SAD are shown in Figure \ref{fig:ages}.
The SAD is shown for a fiducial value of $\alpha = 2000$ and for $\alpha=0$ (no smoothing) and $\alpha=10^5$ to illustrate that the existence and positions of the main peaks are independent of this choice.

\begin{figure*}
\includegraphics[clip,width=0.95\hsize,angle=0,trim=1.5cm 2cm 3cm 3.5cm] {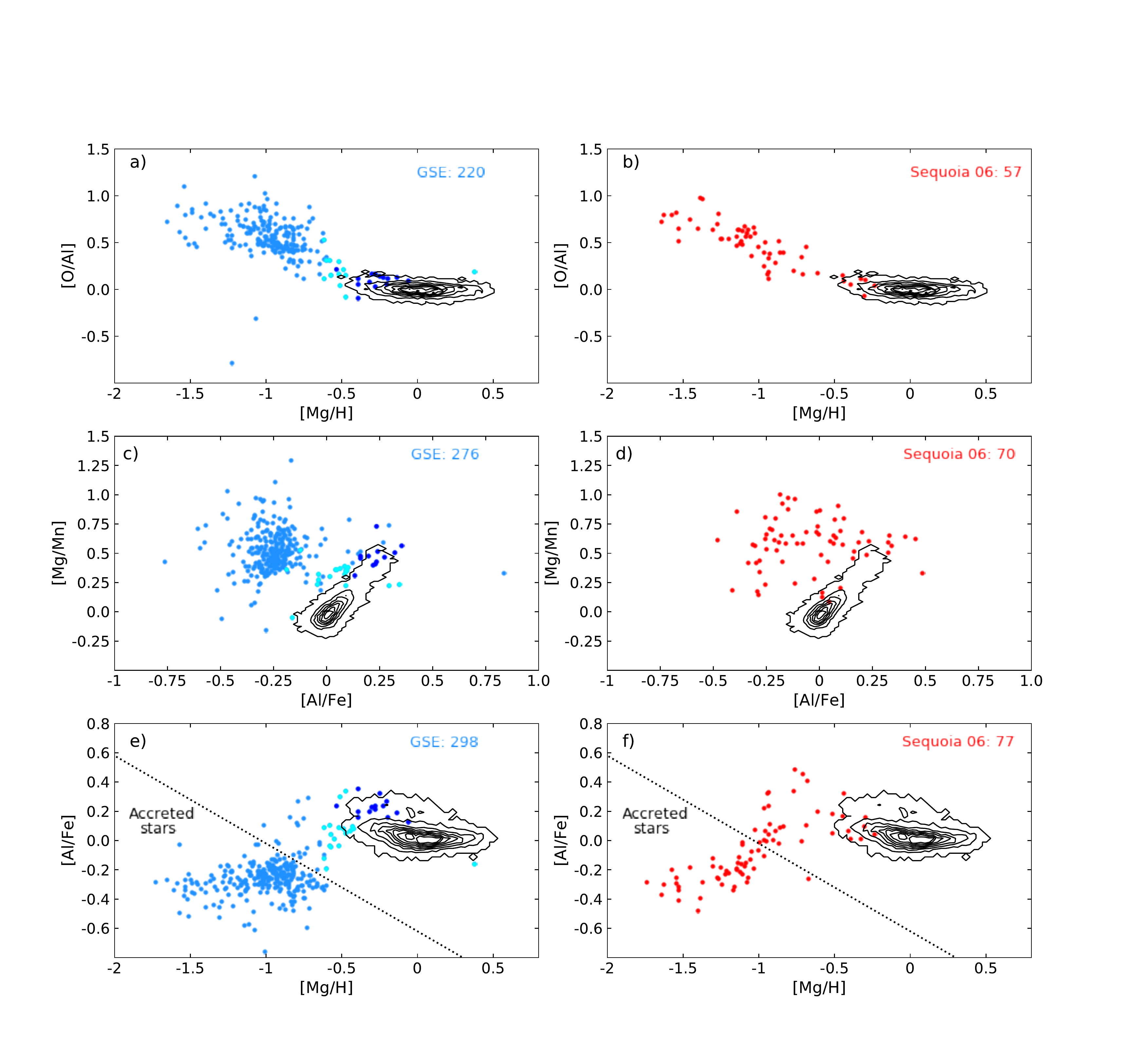}
\caption{APOGEE DR16 + \Gaia DR2 abundance distributions of GSE (blue) and Sequoia (red). The black contours indicate the solar neighborhood distribution. The number of stars shown for each sample is indicated. The GSE stars consistent with the high- and low-[Mg/Fe] sequences of the Milky Way disc, as selected in Figure \ref{fig:MgFe}a, are shown in dark and light blue, respectively.}
\label{fig:abund_SausSeq}
\end{figure*}

We find the GSE stars have an age between $10$ and $13$ Gyr and the age distribution shows a single peak in both individual ages and the SAD. 
For Sequoia 06, the stars are found to have a slightly wider, and on average older, distribution.
The SAD shows two peaks, one between $10$ and $12$ Gyr and the other between $12$ and $14$ Gyr.
These two peaks are most likely associated with {\it in situ} stars and stars from the actual Sequoia population, respectively. The relative contribution of the younger peak is consistent with the {\it in situ} contamination expected from the elemental abundance distributions, see Figures \ref{fig:MgFe} and \ref{fig:abund_SausSeq}.
If true, this indicates that the Sequoia population is slightly older than the GSE population as suggested by \citet{Kruijssen2020}. However, we lack the individual metallicities of the \Gaia DR2 RVS samples in order to firmly establish a connection between the age peaks and the different populations (accreted or formed {\it in situ}).
In any case, the ages confirm that both the GSE and Sequoia populations are old which supports the hypothesis that they could have been accreted some time $\gtrsim 10$ Gyr ago.

Besides the main peaks in the SADs, both GSE and Sequoia have smaller peaks at the oldest age, $\sim20$ Gyr, as well as several younger peaks, less than $8$ Gyr. The $20$ Gyr peak is caused by the isochrone grid edge and is a known feature of age distributions determined using a finite grid of isochrones. The young peaks are likely not all due to uncertainties in the age determinations, but represent the presence of younger stars in the samples \citep[see][for a detailed assessment of the SAD method]{Sahlholdt2021}. In both samples, $\sim10\%$ of the stars have young ages. These could be young stars in the accreted populations or young Milky Way stars. In either case blue straggler stars are the most likely candidate as indicated by the color and magnitude of stars with individual age estimates less than $6$ Gyr, see Figure \ref{fig:CMD}. \citet{Casagrande2020} find the blue straggler fraction in the Milky Way halo, selected as high $V_T$ stars, to be $\sim20\%$. If close binaries are progenitor of blue stragglers, this is consistent with the increasing fraction of close binaries when moving from metal rich to metal poor stars \citet{Moe2019}. Although the young $\alpha$-rich stars seen in the Milky Way disc are not thought to be entirely due to blue straggler stars \citep{Martig2015, Chiappini2015}, \citet{Martig2015} find the young $\alpha$-rich stars comprise approximately $20\%$ of their sample. The kinematics of the young stars do not allow us to distinguish between an accreted or {\it in situ} origin. We also do not have elemental abundances for these \Gaia DR2 RVS stars, therefore their origin is unknown at this time, although [Mg/Fe] or [Al/Fe] measurements would shed light on this.

\begin{figure*}
\includegraphics[clip,width=0.99\hsize,angle=0,trim=2cm 0cm 3cm 1cm] {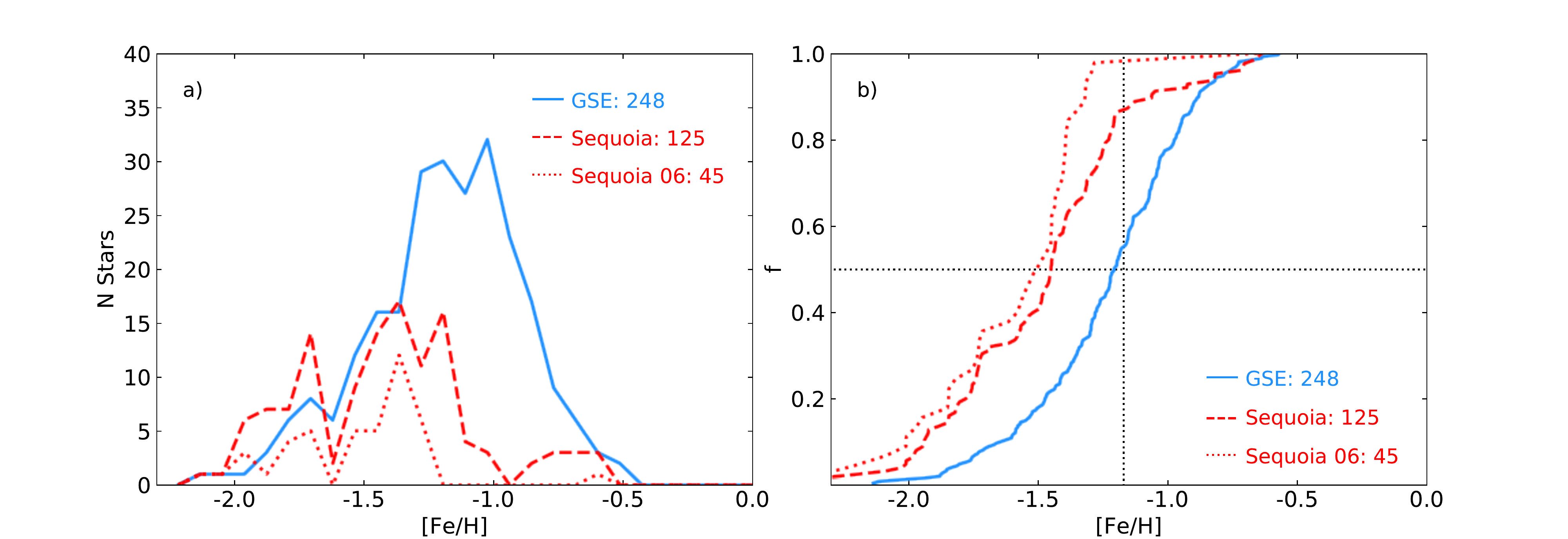}
\caption{Same as Figure \ref{fig:MDF} for the chemically selected members of the GSE (blue solid), Sequoia (red dashed), and Sequoia 06 (red dotted).}
\label{fig:accMDF}
\end{figure*}

\section{Accreted Populations}
\label{sec:accrete}


As noted in Section \ref{sec:abund}, the difference between true GSE and Sequoia stars and the possible {\it in situ} contaminant stars is larger in Al abundance than in Mg. Figure \ref{fig:abund_SausSeq} shows the abundance distribution of the GSE and Sequoia 06 samples in three different abundance spaces that were found empirically to have significant separation of accreted and Milky Way populations. These abundance spaces are [O/Al] vs [Mg/H] (Figure \ref{fig:abund_SausSeq} a, b), [Mg/Mn] vs [Al/Fe] (Figure \ref{fig:abund_SausSeq} c, d), and [Al/Fe] vs [Mg/H] (Figure \ref{fig:abund_SausSeq} e, f). As in Figure \ref{fig:MgFe}, GSE stars consistent with the high- and low-[Mg/Fe] sequences of the Milky Way disc are shown in dark and light blue, respectively. In Figure \ref{fig:abund_SausSeq}, the Sequoia 06 selection is shown, while the abundance distributions of the broader Sequoia selection can be seen in Appendix \ref{ap:seq_comp}. 

GSE stars are significantly separated from the Milky Way disc populations in the [Mg/Mn] vs [Al/Fe] space. They are tightly clustered at high [Mg/Mn] and low [Al/Fe], consistent with the `blob' of likely accreted stars discussed in \citet{Das2020} and consistent with \citet{Hawkins2015}. There are again some stars with abundances consistent with the Milky Way disc populations. The distribution of GSE stars in [Al/Fe] vs [Mg/H] is similar to the distribution in [Al/Fe] vs [Fe/H]. The main population lies at low [Al/Fe] across a $1.0-1.5$ dex range of [Mg/H]. There are high [Al/Fe] stars at high [Mg/H] that appear to be separate from the main population. There a small gap between these groups of stars in [Al/Fe] vs [Mg/H], indicated by the dotted black line in panel e of Figure \ref{fig:abund_SausSeq}. We chose this as our separation criteria between accreted and {\it in situ} stars because of the more distinct separation and the larger number of stars with both Al and Mg measurements (as compared with the smaller number of stars with Mn or O measurements).

The Sequoia 06 stars follow a similar pattern to GSE in [O/Al] vs [Mg/H], overlapping with the disc distribution at high Mg. However, Sequoia 06 is not tightly clustered in [Mg/Mn] vs [Al/Fe] space. As also seen in the [Al/Fe] vs [Mg/H] distribution, Sequoia have a scattered distribution rather than a distinct overdensity. The Sequoia 06 distribution in [Al/Fe] vs [Mg/H] also shows a sequence with little scatter. Unlike GSE, the Sequoia 06 samples does not show an obvious separation in abundance between accreted and {\it in situ} stars. The dotted black line is the same in panels e and f of Figure \ref{fig:abund_SausSeq}.

For this study we chose to use the same elemental abundance-based selection criteria for accreted stars that was derived from the GSE sample and apply it to the Sequoia 06 sample. The abundance patterns of Sequoia depend heavily on the kinematic selection (see Sequoia versus Sequoia 06 samples, Figure \ref{fig:MgFe}), therefore we apply the line defined with the GSE sample and discuss possible repercussions in Section \ref{sec:discussion}.

Figure \ref{fig:accMDF} shows the MDF and CDF of GSE and Sequoia using only stars with [Al/Fe] and [Mg/H] consistent with an accreted origin as indicated in Figure \ref{fig:abund_SausSeq}. Using this selection, GSE has a mean/median [Fe/H] of $-1.23/-1.20$ dex and Sequoia 06 has a mean/median [Fe/H] of $-1.58/-1.49$ dex. The GSE distribution is slightly more metal-poor than in Figure \ref{fig:MDF} and has lost the secondary peak at [Fe/H] $\sim -0.6$. The main peak of the Sequoia 06 sample is more metal-poor than the mean of NGC 3201. Only one star remains at high metallicity, consistent with no peak, but the metal-poor peak retains a few more stars. The Sequoia 06 sample only has 45 stars remaining. Figure \ref{fig:accMDF} also shows the broader Sequoia sample after the chemically accreted constraint has been applied (red dashed line), which contains 125 stars and has a distribution consistent with Sequoia 06.


\section{Discussion}
\label{sec:discussion}

We present MDFs, elemental abundance distributions, and age distributions of the GSE and Sequoia accreted populations using APOGEE DR16 and \Gaia DR2 data. We find that the kinematically selected GSE population has a [Fe/H] peak around $-1.15$ dex and a clean sequence in [Mg/Fe] and [Al/Fe]. In [Mg/Fe] vs [Fe/H], the GSE population shows the classic `$\alpha$-knee' around [Fe/H] $\sim -1.4$, with a high-[Mg/Fe] plateau at $\sim 0.27$ dex. 
The [Al/Fe] vs [Fe/H] sequence is fairly flat at [Al/Fe] $\sim -0.25$, with a possible bump in [Al/Fe] at [Fe/H] $\sim -1.1$.
The GSE population is also strongly clumped in [Mg/Mn] vs [Al/Fe], as observed for accreted stars by \citet{Das2020}. 

We find that the GSE population is consistent with being uniformly old, with a strong peak in age between $10$ and $12$ Gyr. This is in agreement with the mean age of $10$ Gyr found by \citet{Montalban2021} using seven stars with asteroseismic measurements. It is slightly younger (but still consistent within the typical age uncertainties of old stars) than the mean age measured for the blue sequence of the dual halo \citep{Gallart2019, Sahlholdt2019} and the accreted halo stars identified by \citet{Dimatteo2019}, all of which likely contain GSE stars, in addition to other possible accreted populations. \citet{Vincenzo2019} fit a chemical evolution model to GSE members selected by \citet{Helmi2018} from APOGEE and find the [$\alpha$/Fe] vs [Fe/H] distribution and MDF are fit well by a model with a stellar mass of $10^{9.70}$ M$_{\odot}$ and a median age of $12.33$ Gyr.

A few stars are present that likely do not belong to the GSE population but could be heated Milky Way disc or {\it in situ} halo stars. Figures \ref{fig:MgFe} and \ref{fig:abund_SausSeq} highlight these stars, as selected in [Mg/Fe] vs [Fe/H] space, in all abundance spaces. With the exception of two low-[Mg/Fe] stars in Figure \ref{fig:abund_SausSeq} c, these Milky Way stars have elemental abundances consistent with the disc populations.

The [Al/Fe] vs [Mg/H] distribution of GSE has small gap between the low-[Al/Fe], low-[Mg/H] stars and the high-[Al/Fe], high-[Mg/H] stars. We use this gap to define a chemical separation of accreted and {\it in situ} stars. Applying this additional selection criteria to the GSE sample, we find a single peak in the MDF at [Fe/H] $\sim -1.2$, see Figure \ref{fig:accMDF}.

In addition to the possible Milky Way stars identified in [Mg/Fe] vs [Fe/H], there are also some high-[Al/Fe], low-[Fe/H], low-[Mg/H] stars that are not consistent with the main GSE population. As the high [Al/Fe] values at low [Fe/H] are similar to the distribution of NGC 3201, we investigate a possible globular cluster origin of the stars. However, we do not find that these stars have a narrow MDF or an elemental abundance pattern similar to NGC 3201 in other abundance spaces. In particular, the [O/Al] vs [Mg/H] distribution of the high-[Al/Fe], low-[Fe/H] GSE stars is consistent with the rest of the GSE sample. This is not the case for NGC 3201, as can be seen in Figure \ref{fig:seq_comp2}. Additionally, the high-[Al/Fe], low-[Fe/H] GSE stars do not clump in any kinematic space investigated in this work. They are, however, all on retrograde orbits, which suggests they may be a low-[Fe/H] extension of the Milky Way disc.

The Sequoia kinematic selections produce a more complex population. The MDF has a primary peak at [Fe/H] $\sim -1.3$ and a secondary peak at [Fe/H] $\sim -0.7$. The elemental abundance distributions of the larger Sequoia selection show two sequences in both [Mg/Fe] vs [Fe/H] and [Al/Fe] vs [Fe/H], consistent with the Sequoia G1 and G2 populations found by \citet{Monty2020}. One sequence is consistent with the GSE population and one resembles a metal-poor extension of the thick disc. Our more conservative Sequoia 06 selection contains primarily the low [Fe/H], high [Mg/Fe], low [Al/Fe] stars, but still contains some stars consistent with both sequences in the larger Sequoia selection. The Sequoia 06 population is also old with a primary peak in age between $12$ and $14$ Gyr, and a secondary peak between $10$ and $12$ Gyr. 

\citet{Koppelman2019} suggest Sequoia may not be a separate dwarf population due to its large extent in $E_n$ and low [Fe/H]. Instead they argue that the \citet{Myeong2019} Sequoia selection may comprise a different accreted population at low $E_n$ called Thamnos as well as a retrograde, low [Fe/H] extension of GSE at higher $E_n$. Based on APOGEE DR16 elemental abundances, we find that some Sequoia stars have [Mg/Fe] and [Al/Fe] consistent with a low [Fe/H] extension of GSE. This suggests an $L_z$-metallicity gradient in the GSE debris and a radial metallicity gradient in the GSE progenitor, as modeled by \citet{Naidu2021}. This interpretation is also supported by the difference in the elemental abundance patterns between our Sequoia and Sequoia 06 selections, see Figure \ref{fig:MgFe}. The Sequoia selection extends farther towards GSE in kinematic space than Sequoia 06 and also includes more stars with elemental abundances consistent with the GSE population. 

\citet{Naidu2020} consider the highly retrograde halo to host Thamnos at low $E_n$ and three accreted populations at high $E_n$, Sequoia, Arjuna, and I'itoi. Using metallicities from the H3 survey \citep{Conroy2019}, \citet{Naidu2020} find Sequoia has a mean [Fe/H] of $-1.6$ dex, while Arjuna, which contains most of the retrograde high $E_n$ stars, peaks around $-1.2$ dex and I'itoi peaks below $-2.0$ dex. These MDF peaks are not consistent with our Figure \ref{fig:MDF}, although we do find three peaks. The $-1.6$ dex peak in \citet{Naidu2020} is consistent with the low [Fe/H] of Sequoia found by studies using small numbers of high resolution spectra \citep{Monty2020, Aguado2021} and LAMOST \citep{Matsuno2019}. If we reinterpret the Sequoia 06 MDF in Figure \ref{fig:MDF} to match the \cite{Naidu2020} populations, we would associate the main peak at [Fe/H] = $-1.3$ with Arjuna and the metal-poor peak at [Fe/H] = $-1.7$ with Sequoia. This interpretation would suggest that APOGEE DR16 + \Gaia DR2 contains very few Sequoia stars and no I'itoi stars. \citet{Naidu2021} argue that Arjuna is the retrograde debris of GSE, however, we find that the stars contained within the main Fe peak of our Sequoia 06 selection (and most of the Sequoia selection) have a higher [Mg/Fe] than the GSE stars. If we consider the Sequoia selection, instead of Sequoia 06, we find stars at [Fe/H] $> -1.3$ with [Mg/Fe] and [Al/Fe] consistent with the GSE distribution. These stars could comprise the retrograde GSE/Arjuna stars described in \citet{Naidu2021}. It is difficult to separate possible GSE stars from true Sequoia stars below [Fe/H] of $-1.3$.



\subsection{Masses}

Following the method of \citet{Feuillet2020}, which uses the mass-metallicity-redshift relation of \citet{Ma2016} to estimate the progenitor mass of GSE, we estimate the progenitor mass of the Sequoia. As the [Fe/H] of GSE found here is generally consistent with that of \citet{Feuillet2020}, we do not modify the progenitor mass estimate for GSE. The $\sim 0.05$ dex lower [Fe/H] of the chemically selected GSE members does not change the mass estimate from that estimated in \citet{Feuillet2020} within the uncertainties of the [Fe/H] measurements and the mass-metallicity relation. This suggests that mass estimates in the literature based on the metallicity of GSE samples with possible {\it in situ} contamination could also be overestimated. 

\begin{figure}
\includegraphics[clip,width=0.99\hsize,angle=0,trim=0.5cm 0cm 1cm 1cm] {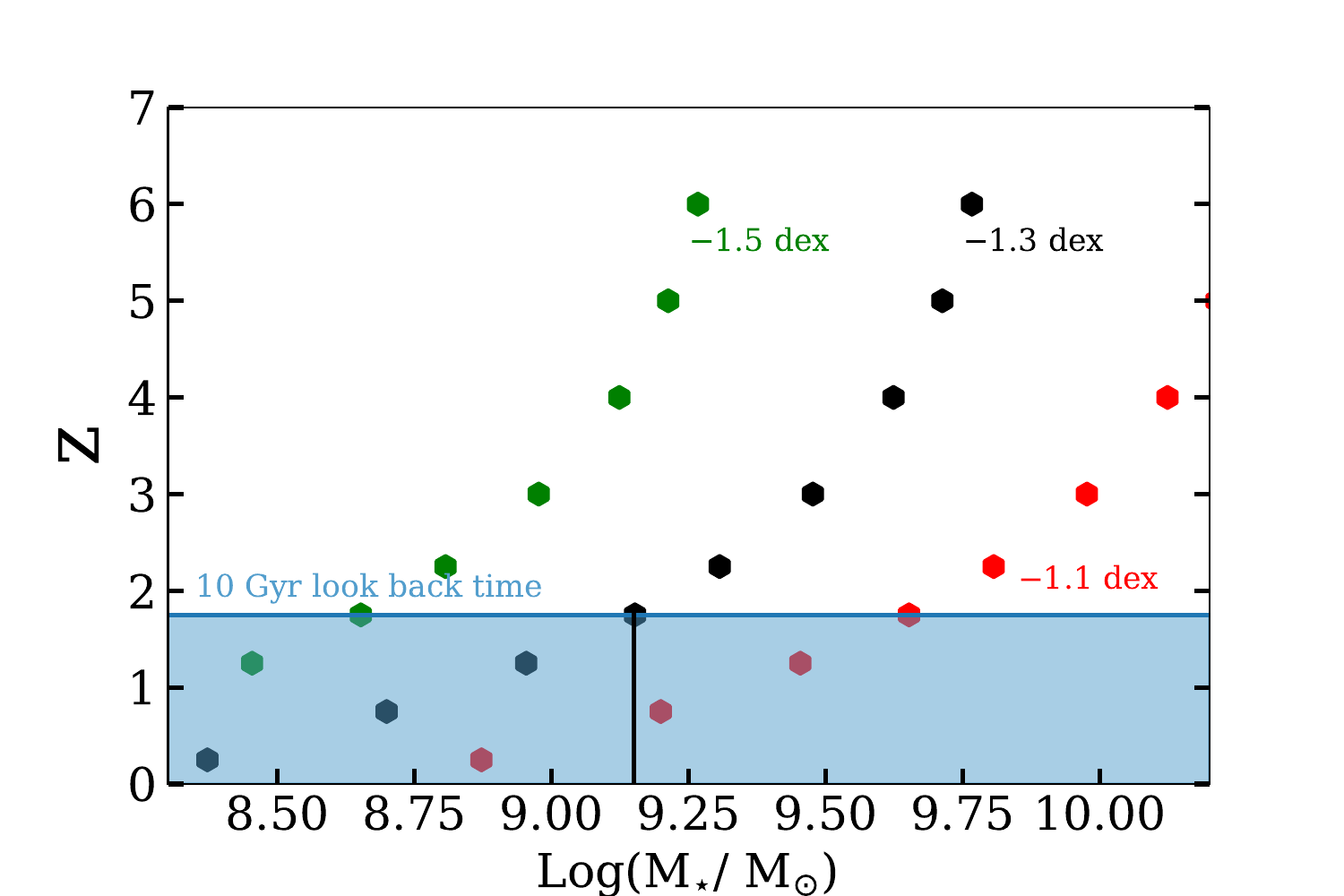}
\caption{The mass-metallicity-redshift relation of \citet{Ma2016}. The shown metallicities correspond to the mean Sequoia [Fe/H] and $\pm 0.2$ dex. }
\label{fig:seqmass}
\end{figure}

Figure \ref{fig:seqmass} shows the redshift versus the progenitor mass relation for [Fe/H] $= -1.5, -1.3, -1.1$ dex. If we assume the Sequoia progenitor was accreted around $Z \sim 1.75$, $10$ Gyr ago, and has [Fe/H] $\sim -1.3$, then the mass is estimated to be $10^{8.65-9.65}$ M$_{\odot}$. If the true [Fe/H] of Sequoia is much lower, as suggested in Section \ref{sec:accrete} where the chemically accreted selection is applied, then the progenitor mass could have been as low as $10^{7.9-8.9}$ M$_{\odot}$. A smaller mass is supported in \citet{Kruijssen2020}, \citet{Myeong2019}, and \citet{Forbes2020}. A  smaller mass is also consistent with a Sequoia metallicity of $\sim -1.3$ dex if it was accreted later, such as $Z \sim 1.5$ as suggested by \citet{Kruijssen2020}.

\subsection{NGC 3201}

NGC 3201 is a globular cluster on a retrograde and highly eccentric orbit, designated a `young halo' cluster by \citet{Mackey2005} and estimated to be $\sim 11$ Gyr by \citet*{Munoz2013}. \citet{Myeong2019} tentatively associate the cluster with the Sequoia accreted population based on metallicity and orbital properties, while \citet*{Massari2019} and \citet{Kruijssen2020} find the progenitor to be either GSE or Sequoia. Several other works associate NGC 3201 with the Gj\"oll Stream found in \Gaia DR2 \citep*{Ibata2019} based on full orbital modelling \citep{Riley2020, Palau2020} and elemental abundance analysis \citep{Hansen2020}. These later works conclude that the cluster is likely the source of the stream, however, a possible connection with the Sequoia progenitor is not discussed. 

The mean metallicity of NGC 3201 is a bit uncertain in the literature. Many studies find a very narrow range of [Fe/H] at around $\sim -1.50$ dex \citep[e.g.][]{Carretta2009, Munoz2013, Simmerer2013, Mucciarelli2015, Magurno2018}. However, \citet{Meszaros2020} found a higher metallicity, $-1.24$ dex, with a scatter in [Fe/H] of $0.1$ dex using APOGEE DR16 data. 

We find the mean [Fe/H] of NGC 3201 members within the Sequoia selection to be $-1.35$ dex with a standard deviation of $0.08$ dex, similar to that found by \citet{Meszaros2020} also using APOGEE DR16 data. This metallicity is consistent with the Sequoia population, see Figure \ref{fig:MDF}. Whether the individual elemental abundances are also consistent with the Sequoia population is a bit more complex, Figures \ref{fig:MgFe}, \ref{fig:seq_comp1}, and \ref{fig:seq_comp2}. 

NGC 3201 has a large range of [Al/Fe] abundance ratios consistent with the Mg-Al anti-correlations common to globular clusters \citep{Meszaros2020}. The Sequoia population also has a rise in [Al/Fe] at the same metallicity as NGC 3201. It is unclear whether the spread in the Sequoia population is due to cluster stars not assigned as members, contamination from {\it in situ} Milky Way stars, or an enrichment in Al within Sequoia field stars. The low [Al/Fe] clump in NGC 3201, designated as the first generation (see \citet{Meszaros2020} for more details), lies just below the Sequoia 06 sequence in [Mg/Fe] and [Al/Fe], see Figures \ref{fig:seq_comp1} and \ref{fig:seq_comp2}. The full NGC 3201 sample also lies lower in [Mg/Mn] than the Sequoia 06 sample.
The second generation stars of NGC 3201, which agree with some Sequoia stars in [Al/Fe] vs [Fe/H], do not agree with the Sequoia 06 [Al/Fe] vs [Mg/H] behavior, shown in Figure \ref{fig:seq_comp2} f, or the Sequoia [O/Al] vs [Mg/H] behavior, Figure \ref{fig:seq_comp2} a. This suggests that there are no stripped NGC 3201 members in the Sequoia field population within the kinematic selection.

Although there are many complexities to account for, we find several inconsistencies in the elemental abundance patterns of NGC 3201 and the Sequoia population.

\subsection{Aluminum as a Diagnostic Element}

Empirically, Al has been found to be depleted in stellar populations of likely accreted populations \citep{Hayes2018}, known accreted populations like GSE and Sequoia (this work), and the recently accreted dwarf galaxy Sagittarius \citep{Hasselquist2017, Hasselquist2019}. However, the cause of the low [Al/Fe] abundances in these low-mass systems is not well-understood. From theoretical models of stellar nucleosynthesis \citep*[e.g.][]{Woosley2002}, Al is thought to be produced primarily in Type II supernovae with a metallicity dependence. This has been confirmed observationally in the Milky Way disc populations \citep{Weinberg2019}. In the GSE sample, we see that [Al/Fe] stays flat with increasing [Fe/H] in the same regime where [Mg/Fe] decreases, see Figure \ref{fig:MgFe} a and b. Unless the Al production increases with increased metallicity at a similar rate to the dilution of Type II supernovae ejecta by Type Ia supernovae, there must be another parameter in the low-mass systems that is causing [Al/Fe] to remain low as [Fe/H] increases and [Mg/Fe] decreases. 


Although Al appears to be a promising diagnostic element for probing stellar populations of low-mass systems, some consideration of the effects caused by an analysis assuming local thermodynamic equilibrium (LTE) is necessary. \citet{Nordlander2017} found that the Al abundance recovered with an analysis assume LTE can differ by up to $1.0$ dex from a non-LTE analysis, depending on the metallicity, effective temperature, and surface gravity of the star in question as well as the specific Al line measured. The NIR lines have a relatively small difference ($\pm 0.1$) in measured [Al/Fe] in giant stars with [Fe/H] between $-1.0$ and $-2.0$ and low [Al/Fe].
We find no dependence of [Al/Fe] on effective temperature or surface gravity within the GSE, Sequoia, or Milky Way disc samples. In addition, our APOGEE samples cover a similar range in stellar parameters, as demonstrated in Figure \ref{fig:CMD}, therefore any non-LTE correction applied to the Al abundances would affect all samples in the same way and the relative abundance patterns would remain. We therefore conclude that our results are not significantly biased by non-LTE effects.

\section{Conclusions}

In this work, we find that the kinematically selected GSE population has Mg and Al abundances consistent with a single accreted system that rapidly formed stars between $10$ and $12$ Gyr ago. This population has a mean metallicity of $-1.2$ dex. The kinematic selection contains a small number of {\it in situ} Milky Way stars identified by their elemental abundances, specifically the {\it in situ} stars have [Mg/Fe] and [Al/Fe] higher than the main GSE population at a given metallicity, consistent with the elemental abundance patterns of the local Milky Way disc. Blue stragglers stars are present within the sample, but cannot be assigned to GSE or the {\it in situ} population given the data used.

The kinematically selected Sequoia population is more complex than GSE. The Mg and Al abundances are indicative of multiple stellar populations occupying the region of the action space map selected, likely consisting of {\it in situ} Milky Way stars, accreted stars, and the globular cluster NGC 3201. We present two kinematic selections of the Sequoia, and a third high energy selection in Appendix \ref{ap:seq_en}. After removal of NGC 3201 members, all selections have signatures of both accreted and {\it in situ} populations. The majority of stars in the Sequoia sample formed between $12$ and $14$ Gyr ago, but a smaller second group of stars is seen with a mean age of around $10$ Gyr. The $10$ Gyr peak is likely comprised of {\it in situ} stars. As in the GSE sample, there are blue stragglers present in the Sequoia sample. 

Sequoia has Mg and Al abundances similar to GSE below [Fe/H] $= -1.3$, but with larger scatter. Above [Fe/H] $= -1.3$, the abundance pattern of the Sequoia sample depends on the specific kinematic selection. The larger Sequoia sample diverges in both [Mg/Fe] and [Al/Fe], while the Sequoia 06 sample contains mainly high [Mg/Fe] and high [Al/Fe] stars. We conclude that the stars most likely to represent the Sequoia population have a mean metallicity between $-1.5$ and $-1.3$ dex, depending on the level of contaminant removal applied, and the stars with [Fe/H] $> -1.0$ are unlikely to be Sequoia members. The small number of Sequoia candidate members makes a robust characterization of the population difficult. However, we conclude that a significant fraction of the stars selected using the action space map are not members, suggesting caution should be used when selecting a box in this space.

To make progress in the efforts to identify and characterize accreted populations found in the Milky Way, we need observations for orders of magnitude more stars in the halo from which elemental abundances can be measured. With high-resolution spectroscopic observations of tens of millions of stars expected in the near future from the WHT Enhanced Area Velocity Explorer \citep[WEAVE,][]{Dalton2018, Dalton2020}, the fifth iteration of the Sloan Digital Sky Survey \citep[SDSS-V,][]{Kollmeier2017}, and the 4-metre Multi-Object Spectroscopic Telescope \citep[4MOST,][]{deJong2019}, the prospects are good. Importantly, these surveys will provide high precision elemental abundance measurements, including Mg and Al, for not only Milky Way halo stars, but also stars in nearby dwarf galaxies. These homogeneous datasets, can be used to better understand the chemical evolution history of accreted populations and examine connections with present-day local dwarf galaxies.

The Dark Energy Spectroscopic Instrument \citep[DESI,][]{2016arXiv161100036D, 2016arXiv161100037D} will, in addition to a large number of galaxies, also observe faint stars in the Milky Way \citep{2020RNAAS...4..188A}. As the stars are faint, $16 < r < 19$, and the survey focuses on high Galactic latitudes, |$b$| $> 22\deg$, a fair number of halo stars will be captured amongst the $\sim$8 million targets in this survey. The spectrographs in DESI have a resolution ranging from about 2000 at the blue end (360\,nm) of the blue arm to 5000 at the red end (980\,nm) of the red arm \citep[][and \url{https://www.desi.lbl.gov/spectrograph/}]{2018SPIE10702E..1FM, 2016arXiv161100037D}. For stellar work this is excellent resolution to obtain radial velocities, but studies performed as preparatory work for WEAVE and 4MOST, as well as the application of the Payne to the stellar spectra of LAMOST \citep{2019ApJS..245...34X} show that more than just radial velocities and a measure of metallicity can be retrieved from such spectra. In particular a measure of Mg as well as Al should be possible to retrieve. Given that DESI focuses on halo areas and goes deep, the impact on the study of the accreted halo should be significant.





\section*{Acknowledgements}

We thank the anonymous referee for their thoughtful and constructive comments.
D.K.F., S.F. and C.L.S. were supported by the grant 2016-03412 from the Swedish Research Council. LC is the recipient of the ARC Future Fellowship FT160100402. Parts of this research were conducted by the ARC Centre of Excellence ASTRO 3D, through project number CE170100013.

This work has made use of data from the European Space Agency (ESA) mission {\it Gaia} (\url{https://www.cosmos.esa.int/gaia}), processed by the {\it Gaia} Data Processing and Analysis Consortium (DPAC, \url{https://www.cosmos.esa.int/web/gaia/dpac/consortium}). Funding for the DPAC has been provided by national institutions, in particular the institutions participating in the {\it Gaia} Multilateral Agreement.

This work uses the following Python packages: {\it galpy}, which can be accessed here: \url{http://github.com/jobovy/galpy}, Astropy \citep{Astropy2013, Astropy2018}.

\section*{Data Availability}

The data underlying this article are available in Gaia Data Release 2 (DOI: 10.1051/0004-6361/201833051) and Sloan Digital Sky Survey Data Release 16 (DOI: 10.3847/1538-4365/ab929e). The datasets were derived from sources in the public domain: \url{https://gea.esac.esa.int/archive/} and \url{https://www.sdss.org/dr16}.

\bibliographystyle{mnras}
\bibliography{saus_seq}

\FloatBarrier

\appendix

\section{APOGEE Fields Removed}
\label{sec:fields}

A list of fields removed from the final APOGEE samples in order to avoid bias from known clusters or accreted populations. The APOGEE targeting parameter `FIELD' was used.

\begin{table}
\caption{APOGEE Fields Removed from Selection of GSE and Sequoia \label{tab:fields}}
\centering
\begin{tabular}{l l l l}
\hline \hline 
\verb|IC342_MGA| & \verb|PAL5-2| & \verb|N1851|      & \verb|N5634SGR2|     \\    
\verb|LMC17| 	 & \verb|M10| 	 & \verb|N2204|      & \verb|N6229| 	    \\	       	   
\verb|LMC7| 	 & \verb|M107| 	 & \verb|N2243-S|    & \verb|SGR1| 	    \\	       	   
\verb|ORIONB-B|  & \verb|M13| 	 & \verb|N2808|      & \verb|SMC5| 	    \\	       	   
\verb|ORPHAN-1|  & \verb|M3| 	 & \verb|NGC3201|    & \verb|TAUL1536| 	    \\     
\verb|ORPHAN-3|  & \verb|M4|  	 & \verb|NGC3201RRL| & \verb|TRIAND-2| 	    \\     
\verb|ORPHAN-5|  & \verb|M55| 	 & \verb|N4147|      & \verb|TRIAND-3| 	    \\     	   
\verb|Omegacen|  & \verb|M92| 	 & \verb|N5466|      & \verb|moving_groups| \\	   
\verb|PAL5-1| 	 & \verb|N1333|  & 		     & 			    \\
\hline
\end{tabular}
\vspace{0cm}
\begin{tablenotes}
\item Note. --- Strings corresponding to the APOGEE parameters `FIELD'.
\end{tablenotes}
\end{table}

\section{Alternate Sequoia Selection}
\label{ap:seq_comp}

This appendix presents the abundance distributions of Sequoia stars when selected using the criteria $J_{\phi}/J_{tot} < -0.4$ and $(J_z - J_R)/J_{tot} < 0.1$. The resulting sample is much larger, but the abundance distributions are more scattered and appear to contain multiple stellar populations.

\begin{figure*}
\includegraphics[clip,width=0.99\hsize,angle=0,trim=1.5cm 0cm 3cm 1cm] {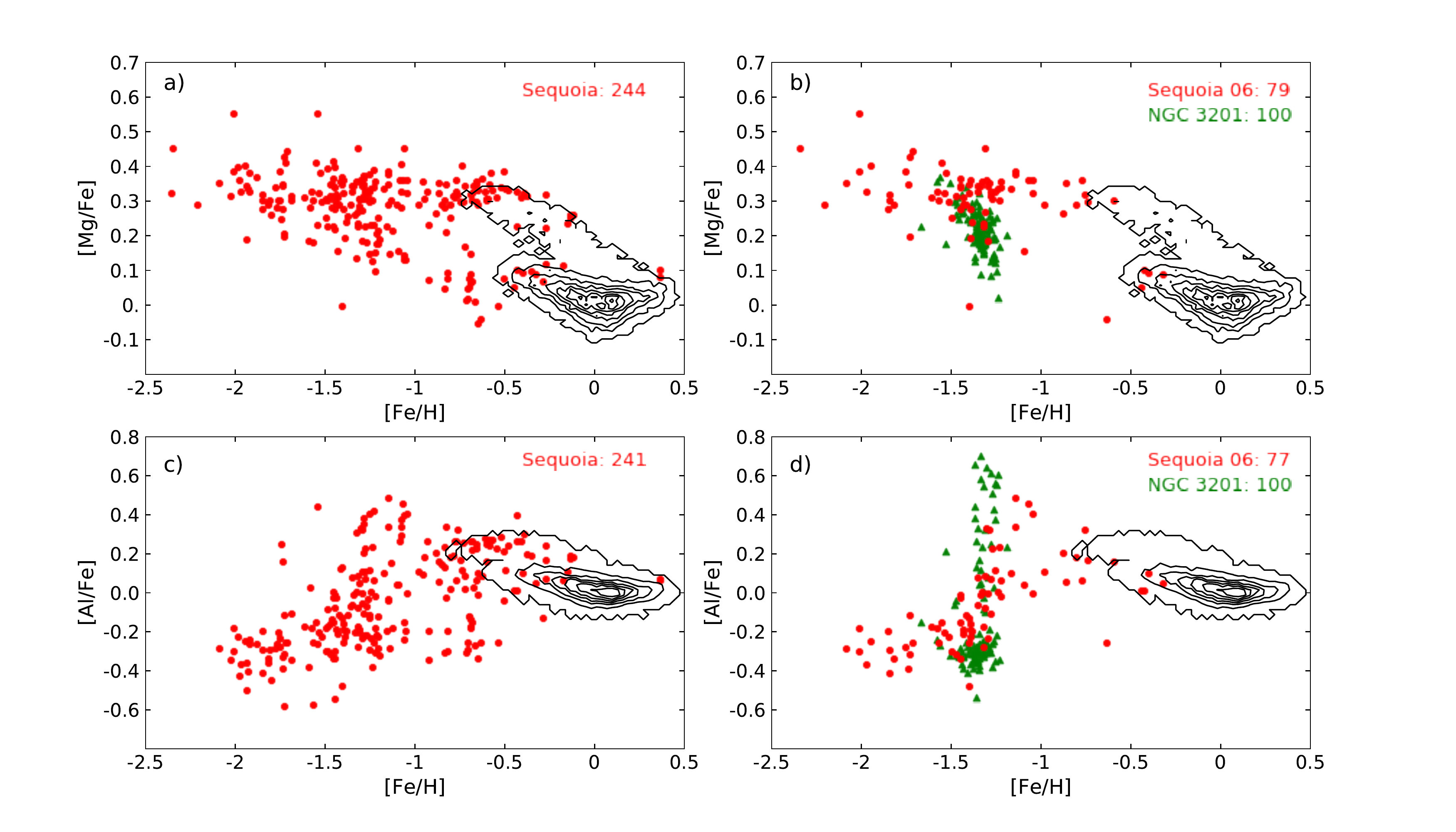}
\caption{A comparison of APOGEE DR16 + \Gaia DR2 abundance distributions of Sequoia using $J_{\phi}/J_{tot} < -0.4$ (left panels) and $J_{\phi}/J_{tot} < -0.6$ (right panels).}
\label{fig:seq_comp1}
\end{figure*}

\begin{figure*}
\includegraphics[clip,width=0.99\hsize,angle=0,trim=1.5cm 2cm 3cm 4cm] {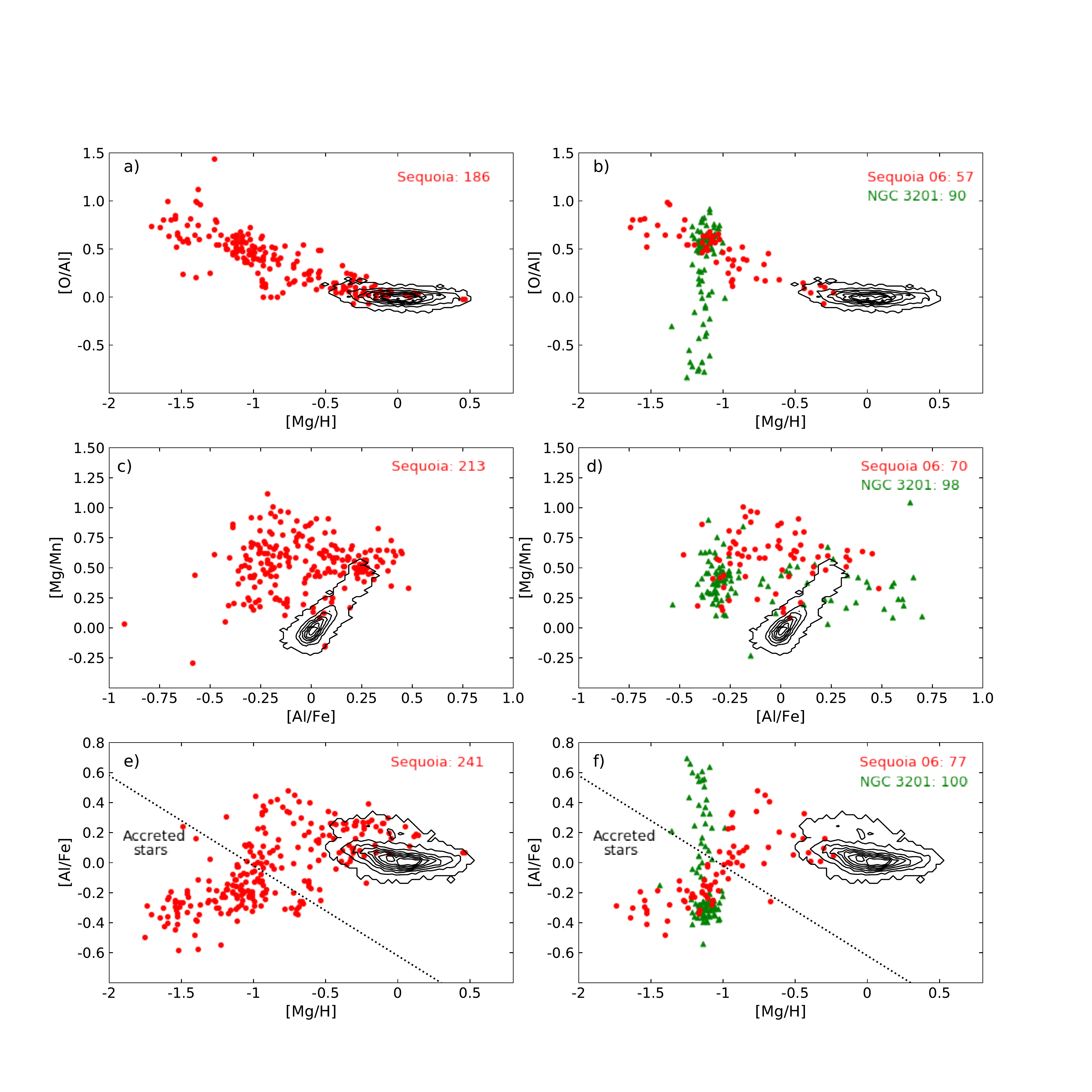}
\caption{A comparison of APOGEE DR16 + \Gaia DR2 abundance distributions of Sequoia using $J_{\phi}/J_{tot} < -0.4$ (left panels) and $J_{\phi}/J_{tot} < -0.6$ (right panels).}
\label{fig:seq_comp2}
\end{figure*}

The first generation stars in NGC 3201 are consistent with the majority of Sequoia stars, but the second generation stars differ from the Sequoia stars with most in these abundance spaces. There are Sequoia stars with high [Al/Fe] at [Fe/H] $\sim -1.3$ (Figure \ref{fig:seq_comp1} c) that are removed in the Sequoia 06 selection, but consistent with the NCG 3201 [Al/Fe] (Figure \ref{fig:seq_comp1} d). This suggests there could be stars not selected as NGC 3201 members, but still within the Sequoia kinematic selection. However, these stars are not consistent with the second generation NGC 3201 stars in other element. While NGC 3201 has a single [Mg/H] abundance and a range of [O/Al], Figure \ref{fig:seq_comp2} b, the Sequoia stars follow a sequence of decreasing [O/Al] with increasing [Mg/H], Figure \ref{fig:seq_comp2} a. This suggests that the Sequoia stars with high [Al/Fe] at [Fe/H] $\sim -1.3$ are not dissolved cluster stars, but likely contaminant Milky Way stars. 

The [Mg/Mn] vs [Al/Fe] (Figure \ref{fig:seq_comp2} c) and [Al/Fe] vs [Mg/H] (Figure \ref{fig:seq_comp2} e) abundance distributions show a large number of stars in the Sequoia selection that have elemental abundances consistent with the Milky Way disc populations. 

\section{High Energy Sequoia Selection}
\label{ap:seq_en}
 
Previous studies typically characterize the Sequoia population as having high orbital energy \citep[e.g.][]{Myeong2019, Naidu2020} while the Thamnos accreted populations occupy the low orbital energy, retrograde kinematic space \citep[e.g.][]{Koppelman2019, Naidu2020}. As seen in Figure \ref{fig:kinematics} g and h, our selection of Sequoia, although based on the \citet{Myeong2019} criteria in the action space map, results in a sample with a large range of orbital energies. This sample also results in a distribution of [Mg/Fe] vs [Fe/H] with both a high- and low-Mg branch, see Figure \ref{fig:MgFe}. Here we investigate potential differences in the orbital energy distributions of Sequoia stars in different elemental abundance regimes. 

\begin{figure*}
\includegraphics[clip,width=0.99\hsize,angle=0,trim=1.5cm 2.3cm 1cm 4cm] {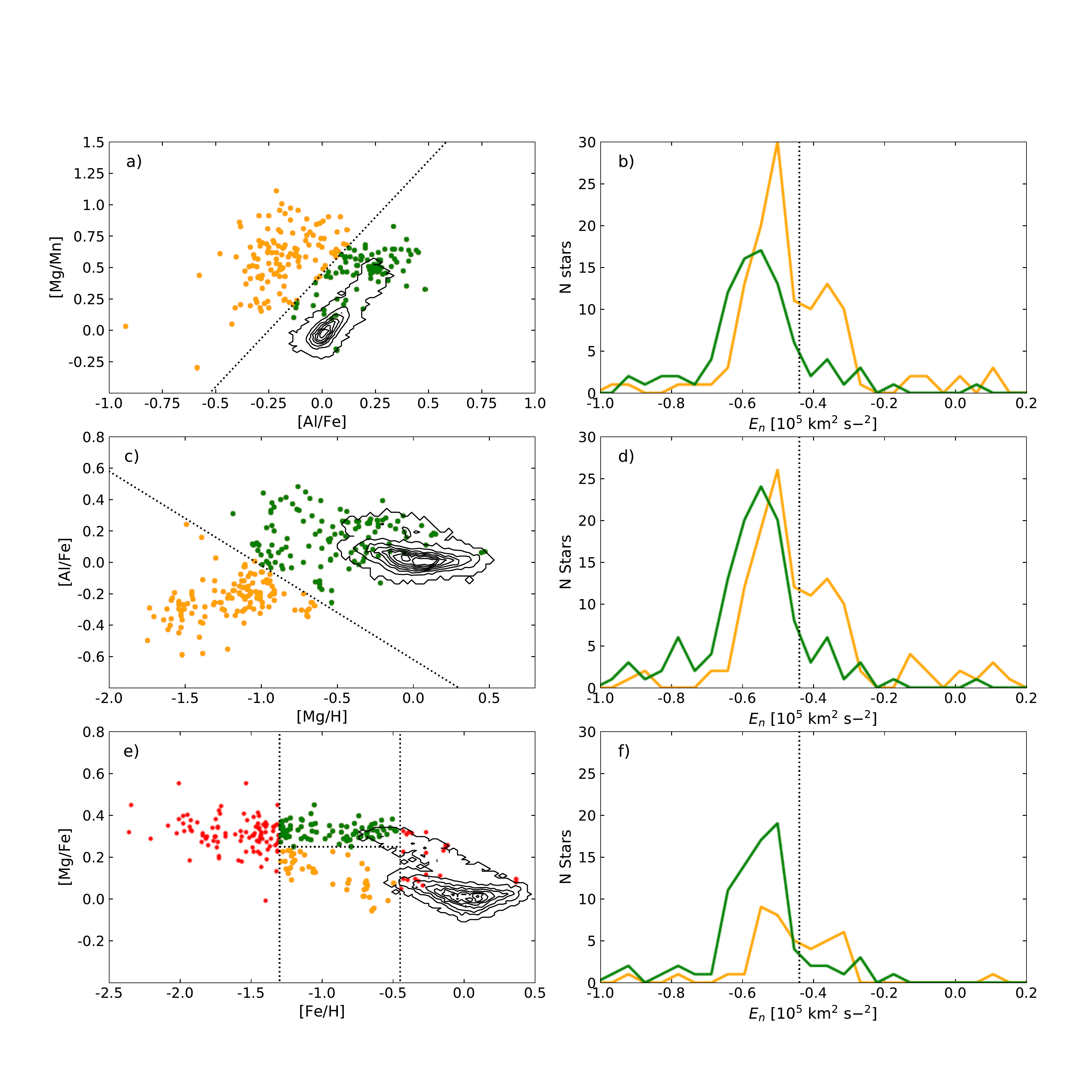}
\caption{The orbital energy distribution of Sequoia subsamples selected based on elemental abundance. Panels a, c, and e show the selection of chemically accreted (yellow) and chemically {\it in situ} (green) stars from the larger Sequoia sample. The edges of the selected regions are shown by the dotted black lines. Stars not selected in either group are shown in red. Panels b, d, and f show the corresponding histograms of the orbital energy of the accreted and {\it in situ} subsamples. The vertical black dotted line indicates the selection of the high energy stars. }
\label{fig:seq_en}
\end{figure*}

Figure \ref{fig:seq_en} shows the orbital energy distributions (b, d, and f) of subsamples of Sequoia selected as chemically accreted (yellow) and chemically {\it in situ} (green) in the [Mg/Mn] vs [Al/Fe] (a), [Al/Fe] vs [Mg/H] (c), and [Mg/Fe] vs [Fe/H] (e) distributions. The edges of the selection regions are indicated by the black dotted line. The corresponding energy histograms show that although the two subsamples have a similar peak energy, the chemically accreted subsamples (yellow) have an excess of high energy stars. This high energy excess is present in all three elemental abundance spaces investigated. 

\begin{figure*}
\includegraphics[clip,width=0.99\hsize,angle=0,trim=1.5cm 0.5cm 1cm 1cm] {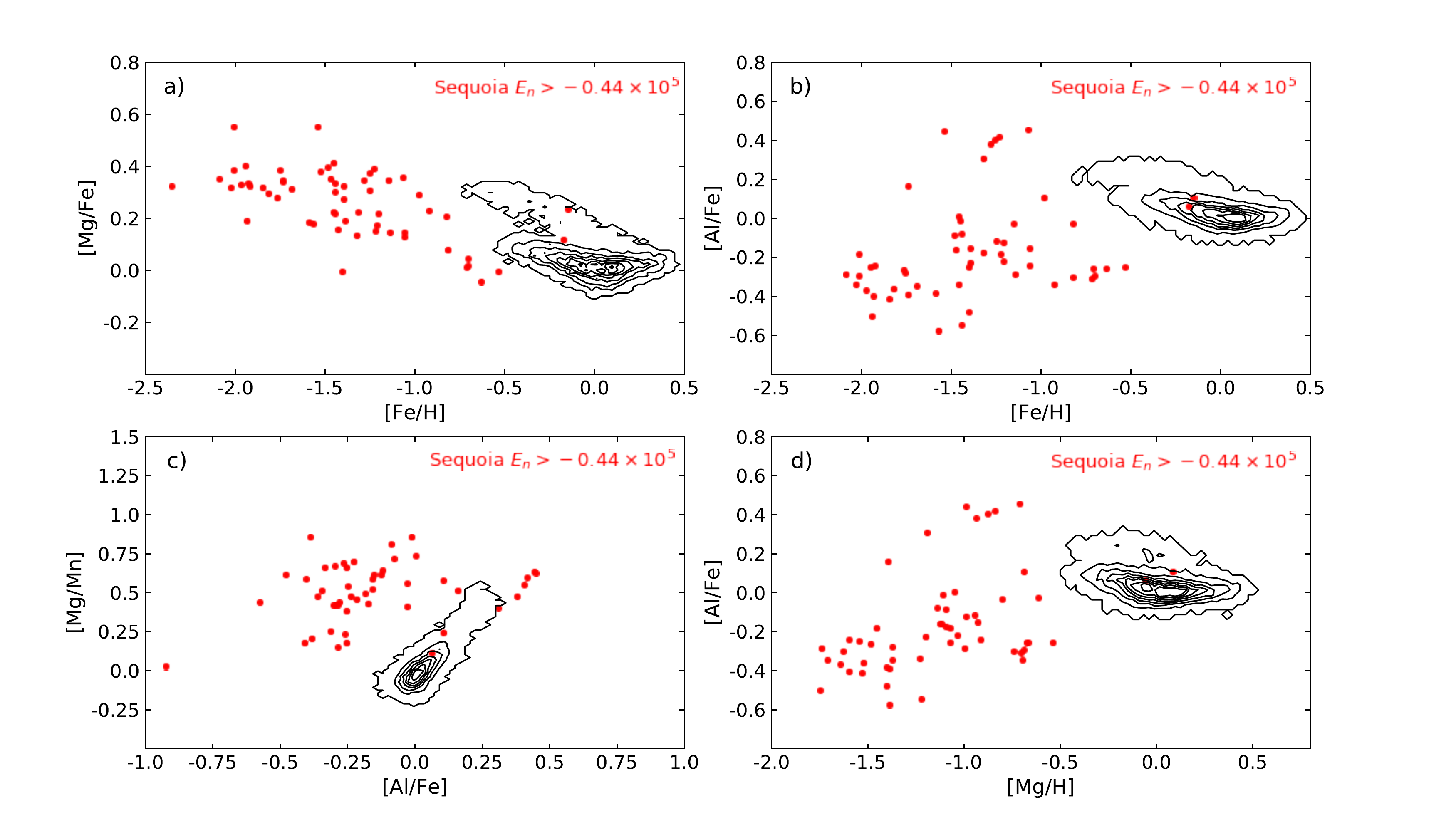}
\caption{The elemental abundance distribution of Sequoia stars selected to be high energy as indicated in Figure \ref{fig:seq_en}. The black contours indicate the solar neighborhood distribution. }
\label{fig:seq_en_abund}
\end{figure*}

Assuming Sequoia is in fact a high energy population, we define an energy limit for Sequoia members ($E_n > -0.44 \times 10^5$ km$^2$ s$^{-2}$) based on the minimum between the main energy peak and the high energy excess in the yellow histograms. Figure \ref{fig:seq_en_abund} show the distribution of this high energy Sequoia sample in four elemental abundance spaces. As compared to the full Sequoia and Sequoia 06 distributions, see Figures \ref{fig:seq_comp1} and \ref{fig:seq_comp2}, the high energy Sequoia do not show as much evidence of two populations as the full Sequoia selection (i.e. $J_{\phi}/J_{tot} < -0.4$), however, the comparison with the Sequoia 06 sample is less clear. 

The [Mg/Fe] vs [Fe/H] distribution of the high energy Sequoia is consistent with a single low-Mg sequence, while the Sequoia 06 distribution retains evidence of multiple sequences. The [Al/Fe] vs [Fe/H] and [Al/Fe] vs [Mg/H] distributions of Sequoia 06 suggest a single population sequence, while the high energy Sequoia distributions have a second high-Al sequence. The [Mg/Mn] vs [Al/Fe] distributions of the Sequoia 06 and high energy Sequoia selections are similarly scattered and both shown signs of {\it in situ} contamination.

Although we do not use an energy selection in our main analysis, we find this exercise results in an interesting suggestion of multiple populations being present within the action space map selection of the Sequoia. The chemically accreted samples in Figure \ref{fig:seq_en} all have a primary peak at low energy, only slightly higher than the energy peak of the chemically {\it in situ} samples. This bimodality of the chemically accreted samples suggests either 1) there is another, low energy accreted population, such as Thamnos, occupying the sample region of the action space map as Sequoia, 2) Sequoia has a bimodal energy distribution, or 3) there is a significant contribution of {\it in situ} stars to the regions of elemental abundance space thought to comprise accreted populations.

\bsp
\label{lastpage}
\end{document}